\documentclass[journal]{IEEEtran}
\IEEEoverridecommandlockouts

\usepackage{cite}
\usepackage{amsmath,amssymb,amsfonts}
\usepackage{algorithm}
\usepackage{graphicx}
\usepackage{textcomp}
\usepackage{xcolor}
\usepackage[hidelinks,bookmarks=false]{hyperref}
\usepackage[caption=false,font=footnotesize]{subfig}
\usepackage{comment}
\usepackage{tikz}
\usepackage{varwidth}
\usepackage{enumitem}
\usepackage{algpseudocode}
\usepackage{float}
\usepackage{amsthm}
\usepackage{multirow} 
\usepackage{placeins}

\usetikzlibrary{arrows.meta, positioning, shapes.geometric}
\newcommand{\Rnum}[1]{\uppercase\expandafter{\romannumeral #1\relax}}

\def\BibTeX{{\rm B\kern-.05em{\sc i\kern-.025em b}\kern-.08em
    T\kern-.1667em\lower.7ex\hbox{E}\kern-.125emX}}

\begin{document}
\bstctlcite{BSTcontrol}
\title{
VitalFR3: Wireless Vital Sign Sensing in the \\ Upper Mid-Band (FR3)







}

\author{
\IEEEauthorblockN{Gholamreza Bakhshi and Hatem Abou-Zeid}\\
\IEEEauthorblockA{\textit{Dept. of Electrical and Software Engineering, WAVES LAB, University of Calgary, Canada} \\
\{gholamreza.bakhshi, hatem.abouzeid \}@ucalgary.ca}
\thanks{This work was supported by infrastructure funded by the Canada Foundation for Innovation (CFI) John R. Evans Leaders Fund (JELF) and the Government of Alberta, Research Capacity Program (RCP).}
}

\maketitle

\begin{abstract}

Frequency range 3 (FR3) is being explored for sixth-generation (6G)
integrated sensing and communication (ISAC). However, experimental evidence
on the feasibility of physiological sensing in this band is limited. This paper presents VitalFR3, an SDR-based
frequency-modulated continuous-wave (FMCW) measurement platform that
integrates a direct-sampling Ettus USRP X440 with a SignalCraft SC2470
frequency converter. To the best of our knowledge, this is the first
calibrated SDR-based FMCW platform evaluated for respiration and
cardiac-rate estimation from human measurements across multiple FR3
center frequencies. Using a $100~\mathrm{MHz}$ waveform, we operate the
platform at six RF center frequencies from $10$ to $22.5~\mathrm{GHz}$.
We demonstrate how to characterize and calibrate the complete VitalFR3 transmit and receive
path directly from loopback recordings, achieving a magnitude error
below $0.061~\mathrm{dB}$ at all six frequencies.
We then propose a low-complexity end-to-end processing pipeline for vital sign estimation.
This involves the identification of a persistent target-associated response in fast time, respiratory tracking in slow time, followed by
cardiac-rate estimation. 
The system is evaluated across 16 human measurements, achieving a respiratory-rate
mean absolute error of $0.758$ breaths per minute,
and the majority of the cardiac-rate estimates fall within $\pm5~\mathrm{BPM}$ of
their reference rates. In a separate breath-hold measurement, the
estimated cardiac rate is $84.3~\mathrm{BPM}$ versus approximately
$84.5~\mathrm{BPM}$ from the reference. These results establish VitalFR3
as a calibrated and configurable measurement platform for experimental
physiological sensing across FR3.

\end{abstract}


\section{Introduction}
Sixth-generation (6G) wireless systems are expected to integrate
communication and sensing through Integrated Sensing and Communication
(ISAC), enabling shared radio resources to support both data transmission
and environmental sensing~\cite{11519592,
pegoraro2025toward_multiband_sensing_fr3,11573792}. Contactless vital sign sensing
is one potential application, since radio signals can capture the small
body-surface motions caused by respiration and cardiac activity without requiring wearable
sensors~\cite{11432470}. Frequency range 3
(FR3), spanning 7.125--24.25~GHz, lies between sub-6~GHz and millimeter-wave (mmWave)
bands, and offers a distinct combination of propagation characteristics, available bandwidth, and phase sensitivity to the small displacements associated with vital signs. While these properties create opportunities for physiological sensing, there are also unique propagation challenges associated with the FR3 band.



Indoor FR3 channels can exhibit rich multipath and considerable delay
spread, with waveguide-like propagation reported in some indoor
environments~\cite{shakya2024comprehensive_fr1c_fr3}. These conditions
matter for vital sign sensing because the small phase variations caused by
respiration and cardiac motion must be measured in the presence of multiple
propagation paths. Most experimental FR3 studies have so far focused on
propagation measurements, channel characterization, communication
testbeds, and macroscopic sensing tasks~\cite{ying2025upper_mid_band_factory,
shakya2024propagation_indoor_fr1c_fr3,
11432645,
pegoraro2025toward_multiband_sensing_fr3,
bomfin2025multi_band_sensing}. However, experimental studies of respiration and
cardiac sensing in this band remain limited. This leaves an open question
of how a programmable FR3 measurement system can be built and utilized for
physiological sensing.

Physiological sensing in FR3 requires a measurement platform that can
generate and acquire a controlled waveform while preserving the small
phase variations associated with respiration and cardiac motion. Compared
with more established mmWave sensing bands, FR3 has a less mature
ecosystem of integrated sensing instrumentation. Experimental work in
this band therefore requires not only sensing methods, but also the
development and characterization of the measurement platform itself.
To meet this need, we develop VitalFR3, a programmable FR3 measurement
platform that combines an Ettus USRP X440 software-defined radio (SDR) with
a SignalCraft SC2470 frequency up/down converter. The X440 direct sampling
architecture provides control of waveform generation, sampling, and
acquisition, while the SC2470 provides frequency translation between the
X440 intermediate frequency (IF) and the selected FR3 RF center frequency. 
The VitalFR3 architecture supports instantaneous bandwidths up to
$1.6~\mathrm{GHz}$, enabled by the wideband capabilities of the X440 and
SC2470.

While the X440 enables IF phase coherence and large bandwidths, it also introduces two instrumentation challenges. First, vital
sign sensing requires stable phase measurements together with direct
control of the transmitted waveform, sampling, and acquisition timing.
The X440 uses a direct-sampling architecture in which the
balun-coupled front ends are connected directly to the data converters
~\cite{10643510}, which means that it does not use internal analog mixers
or local oscillators for frequency conversion, nor analog filtering in the
analog signal path. Therefore, the converter rate, sampling rate, and placement
of the transmitted waveform within the appropriate Nyquist zone must be
explicitly controlled.
Second, stable phase measurements from the X440 alone do not characterize
the complete sensing path. The SC2470 and fixed components also affect
the amplitude, phase, and timing of the recorded signal - hence the complete
transmit and receive path must be characterized to establish a consistent
measurement reference.

Beyond these instrumentation challenges, physiological recovery and
validation introduce further requirements. Calibrated FR3 measurements do
not directly provide physiological information. In indoor FR3 environments,
rich multipath and delay dispersion can complicate target localization and
the recovery of the small phase variations associated with respiration and
cardiac motion. An advanced signal processing pipeline is therefore needed to convert the
calibrated complex I/Q measurements into physiological estimates. 
Cardiac sensing
also faces a challenging ambiguity when respiratory harmonics enter the cardiac
frequency range. In this case, frequency alone may not be sufficient to
distinguish the physiological origin of the observed component.

To the best of our knowledge, this paper is the first experimental
study to develop and evaluate an SDR-based FR3 platform
for contactless respiration and heart-rate estimation across multiple FR3 center frequencies.
The main contributions of this work are summarized as follows:
\begin{itemize}
\item \textbf{Programmable FR3 Measurement Platform:}
We introduce VitalFR3, a programmable FR3 measurement platform established by integrating the X440 direct-sampling SDR
with the SC2470 FR3 frequency converter. We describe the acquisition configuration needed to correctly
control the converter rate, sampling rate, waveform placement, and
timing in this architecture.
The experimental configuration uses
a 100-MHz FMCW waveform at six RF center frequencies between
10 and 22.5~GHz under a fixed acquisition plan. This platform is scalable to substantially wider bandwidths and provides a basis for examining physiological measurements and other FR3 sensing use-cases using direct sampling.

\item \textbf{End-to-End Calibration of the Platform:}
We demonstrate how to characterize and calibrate the complete VitalFR3 transmit and receive
path directly from loopback recordings. This involves deriving the required corrections
and timing reference for each RF center frequency. Across all six RF
center frequencies, our calibration procedure reduces the remaining magnitude error
to below $0.061~\mathrm{dB}$ and establishes the measurement reference
used by the subsequent sensing pipeline.

\item \textbf{Low-Complexity Vital Sign Processing Pipeline:}
We propose a low-complexity end-to-end processing pipeline for vital sign estimation.
This involves the identification of a persistent target-associated response in fast time, respiratory tracking in slow time, followed by
cardiac-rate estimation. A low-complexity frequency-profile
estimator is used for both fast-time response selection and
slow-time respiratory tracking without covariance estimation or
eigen decomposition. The resulting target-centric phase measurements are then used for cardiac
rate estimation. 

\item \textbf{Experimental Physiological Validation:}
We experimentally validate VitalFR3
using 16 human measurements,
together with an end-expiratory apnea measurement. 
Our results demonstrate excellent agreement with the 
reference respiratory rates achieving a mean absolute error (MAE) of $0.758$ breaths
per minute ($\mathrm{BrPM}$) and  the majority of the cardiac estimates are within
$\pm5~\mathrm{BPM}$ of the cardiac reference. The fine cardiac estimation was challenging
in some measurements due to the presence of the respiratory signal's harmonics within the cardiac frequency range.
We systematically characterize this respiratory harmonic ambiguity in the FR3 measurements using a fixed frequency criterion together with
time--frequency inspection. 
Our analysis shows that respiratory
harmonic proximity is associated with several of the larger cardiac
estimation errors and quantifying it provides a confidence measure of the algorithm's results.
The apnea measurement achieves a $0.2~\mathrm{BPM}$ MAE, providing additional evidence that cardiac-related motion is better recovered when respiratory harmonics are reduced.



\end{itemize}

The remainder of this paper is organized as follows.
Section~\ref{Related_Work} reviews the relevant literature on FR3
measurement and radar-based physiological sensing.
Section~\ref{sec:platform} describes the VitalFR3 measurement platform,
waveform and acquisition design, and end-to-end calibration.
Section~\ref{sec:baseline_processing} presents the processing pipeline
applied to the calibrated measurements.
Section~\ref{sec:experimental_validation} presents the experimental
physiological validation, including the baseline evaluation, apnea
positive control, and respiratory-harmonic ambiguity assessment.
Finally, Section~\ref{sec:conclusion} concludes the paper.

\section{Related Work}\label{Related_Work}
Contactless vital sign sensing has been demonstrated using a range of RF
platforms and operating frequencies. At sub-6~GHz, WiFi-based systems have
used wireless channel measurements for respiration monitoring
~\cite{11432470,Andreas}. Radar-based systems have also been used for
respiratory and cardiac sensing at higher frequencies, including
$24~\mathrm{GHz}$ Doppler radar~\cite{11622583}, $60~\mathrm{GHz}$
frequency-modulated continuous-wave (FMCW) radar~\cite{10579790}, and
$77~\mathrm{GHz}$ FMCW radar~\cite{11418708}. These radar systems measure
the small chest motions associated with respiration and heartbeat through
changes in the received signal phase. FMCW radar platforms can also use range
information to locate the target before extracting the physiological
motion~\cite{10579790,10909288}. Together, these studies show that
contactless respiration and cardiac sensing can be performed across
different RF bands and sensing architectures.

Recent work has also examined how practical measurement conditions affect
radar-based vital sign sensing. Body motion can distort the phase variations
used for respiratory and cardiac estimation, and several studies have
developed methods for vital sign estimation in the presence of random or
nonperiodic motion~\cite{10909288,10648788}. Other work has combined multiple range bins and 
channels to improve cardiopulmonary signal recovery~\cite{11418708}.
Robustness has also been studied under speaking conditions,
where respiratory motion becomes less regular and the weaker cardiac
component is more difficult to recover~\cite{11080405}. In addition to signal
processing, front-end configuration can affect measurement quality; for example,
polarization mismatch has been shown to influence cardiac sensing accuracy in
a $24~\mathrm{GHz}$ Doppler radar system~\cite{11622583}. These studies show
that reliable vital sign sensing depends on both physiological signal
processing and the conditions under which the radar measurements are
acquired.
Prior studies relevant to FR3 have so far focused mainly on
propagation, channel characterization, and, more recently, target sensing.
Comparative ray-tracing studies have also examined FR3 propagation and
coverage across suburban, urban, and high-rise urban environments
~\cite{11577293}, including cellular links in dense high-rise scenarios
~\cite{11587034}.
Xu et al. developed a channel model for FR3 extremely large-scale
multiple-input multiple-output (XL-MIMO) systems that accounts for
near-field propagation and spatial non-stationarity~\cite{11327450}.
Indoor measurements at $6.75$ and $16.95~\mathrm{GHz}$ have characterized
path loss and delay spread in office/laboratory and factory environments
~\cite{shakya2024comprehensive_fr1c_fr3,
ying2025upper_mid_band_factory}. Material penetration loss has also been
measured across these frequencies
~\cite{shakya2024comprehensive_fr1c_fr3}.
Bomfin et al. examined multiband channel sensing at
$6.5$, $8.75$, $10$, and $15~\mathrm{GHz}$ and evaluated the
reliability of estimating channel components
~\cite{bomfin2025multi_band_sensing}.
Moving from channel characterization
toward target sensing, Pegoraro et al. studied coherent multiband ranging of object and human targets over $6$--$22~\mathrm{GHz}$ ~\cite{pegoraro2025toward_multiband_sensing_fr3}. 
At the network level,
Shankar et al. investigated multistatic sensing in FR3, using
geometry-aware multi-link fusion for target localization and velocity
estimation~\cite{11550061}.
Nieman et al. developed
a reconfigurable FR3 channel sounder that supports monostatic and
bistatic measurements and includes calibration of the transmit and receive
measurement paths~\cite{11432645}. 
Experimental physiological sensing across FR3 remains
limited, particularly for programmable measurement platforms that
combine system characterization and calibration with respiratory and
cardiac sensing across multiple FR3 center frequencies.

\section{VitalFR3 Measurement Platform, Waveform Design, and Calibration}
\label{sec:platform}
\begin{figure*}[!t]
    \centering
    \includegraphics[width=\textwidth]{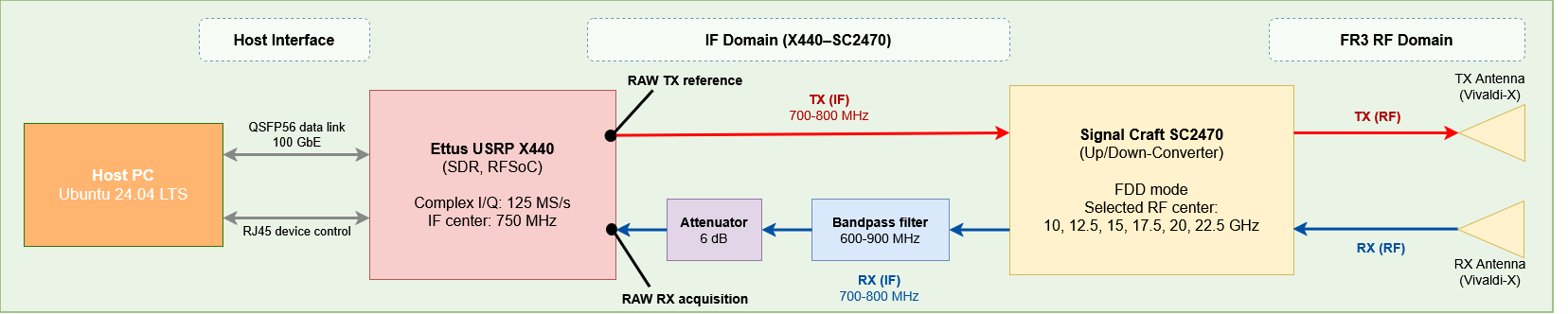}
    \caption{End-to-end VitalFR3 hardware architecture and signal path.}
    \label{fig:FR3_system_schematic}
\end{figure*}
VitalFR3 is implemented as a programmable FR3 measurement platform built
around an Ettus USRP X440 and a SignalCraft SC2470 up/down converter.
The X440 provides waveform generation and acquisition, while the SC2470
translates the signal to and from the selected FR3 center frequency.
Because the recorded data contain the response and timing of the complete
configured measurement path, this path is characterized and calibrated
before the data are passed to the vital sign processing workflow of
Section~\ref{sec:baseline_processing}.

\subsection{FR3 Measurement Architecture and Signal Path}
\begin{figure}[t]
    \centering
    \subfloat[]{%
        \includegraphics[scale=0.0255]{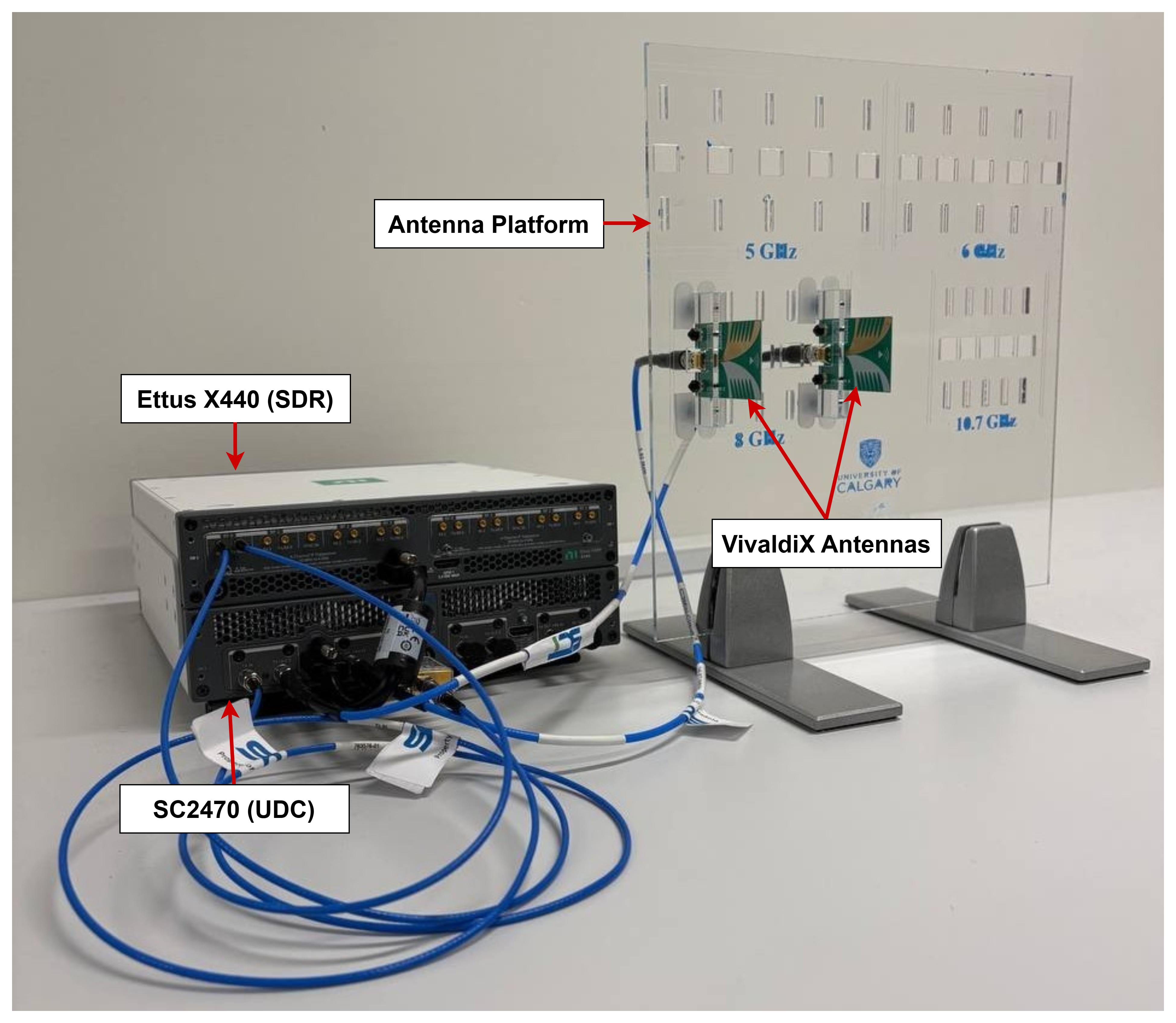}
        \label{fig:FR3_Hardware}
    }
    \subfloat[]{%
        \includegraphics[scale=0.14]{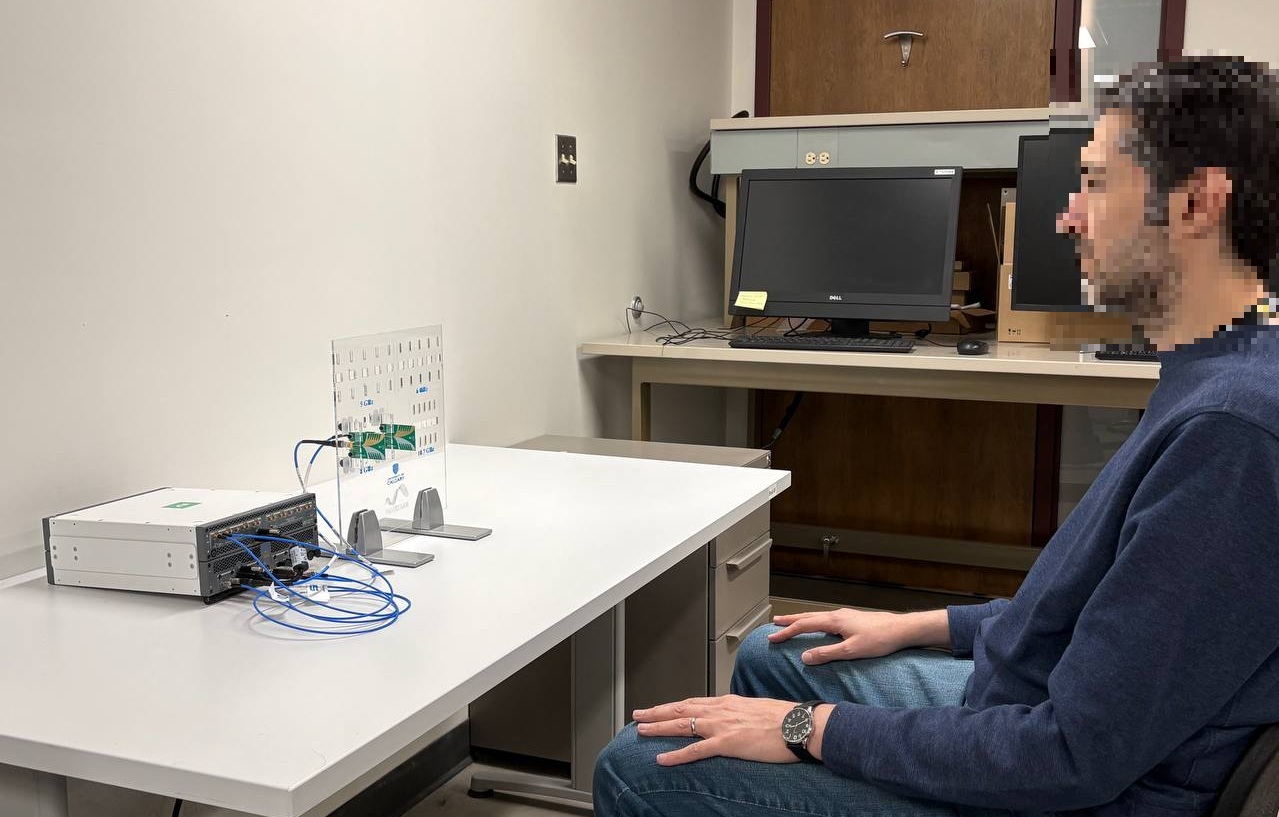}
        \label{fig:Person_of_Interest}
    }
    \caption{VitalFR3 measurement platform:
    (a) hardware implementation comprising the Ettus USRP X440,
    SignalCraft SC2470, and fixed Vivaldi antenna assembly;
    (b) geometry used for the human measurements with the seated participant
    positioned approximately $1~\mathrm{m}$ in front of the antenna
    assembly.}
    \label{fig:FR3_Platform}
\end{figure}

Figures~\ref{fig:FR3_system_schematic} and \ref{fig:FR3_Hardware}
show the end-to-end VitalFR3 measurement chain and the corresponding
physical implementation, respectively. 
The X440 complex in-phase/quadrature (I/Q) interface is operated at a
sampling rate of $125~\mathrm{MS/s}$ with a $1~\mathrm{GHz}$ conversion
rate. The X440--SC2470 interface is centered at an IF of $750~\mathrm{MHz}$ in the VitalFR3 configuration.
The SC2470 translates the IF signal to and from one selected FR3 RF center
at a time. 
The platform architecture supports instantaneous bandwidths up to
$1.6~\mathrm{GHz}$, while the measurements reported in this work use
the fixed $100~\mathrm{MHz}$ waveform configuration described below.
The measurement campaign uses RF center frequencies of
$10$, $12.5$, $15$, $17.5$, $20$, and $22.5~\mathrm{GHz}$, with the 
same X440 sample plan and fixed receive path configuration
retained for the calibration and radar measurements at each RF center.

The SC2470 is connected to separate transmit and receive Vivaldi-X
ultra-wideband antennas mounted on a rigid custom acrylic fixture to
maintain a fixed transmit--receive geometry throughout the measurements.
On reception, the FR3 return is down-converted by the SC2470 to the IF
domain and passes through the Mini-Circuits ZX75BP-750-S+
$600$--$900~\mathrm{MHz}$ bandpass filter and a $6$-dB attenuator
before reaching the X440 receive interface. Both components form part of the fixed receive path.
For the human measurements,
the seated participant was positioned approximately $1~\mathrm{m}$ in
front of the antenna assembly, as shown in
Fig.~\ref{fig:Person_of_Interest}.

Continuous I/Q samples are streamed between the X440 and a Linux host
workstation through a $100~\mathrm{GbE}$ QSFP56 connection using a
dual-QSFP56 Mellanox ConnectX-6 Dx network interface, while a separate
RJ45 link is used for device control. The acquisition workstation runs
Ubuntu 24.04 LTS and is equipped with a $5.4~\mathrm{GHz}$ Threadripper
PRO 9975WX processor and $128~\mathrm{GB}$ of ECC DDR5-5600 memory.

\subsection{FMCW Waveform and Acquisition Design}
VitalFR3 employs a linear up-chirp FMCW waveform with a designed
$100~\mathrm{MHz}$ sweep bandwidth. At the X440 complex I/Q interface,
the active chirp is designed to span $-50$ to $+50~\mathrm{MHz}$
relative to the $750~\mathrm{MHz}$ IF center, corresponding to a
$700$--$800~\mathrm{MHz}$ IF sweep. The SC2470 translates this waveform
to the selected FR3 RF center, and the realized waveform is verified
directly from the recorded transmit samples later in this subsection.

The fixed waveform sample plan comprises $65{,}536$ active samples and
$16{,}384$ idle samples per chirp, corresponding to a
$524.288~\mu\mathrm{s}$ active sweep and a chirp repetition interval
(CRI) of $655.36~\mu\mathrm{s}$. Each frame contains 15 consecutive
chirps and spans $9.8304~\mathrm{ms}$. The complete waveform and
acquisition timing parameters are summarized in
Table~\ref{tab:waveform_configuration}.

\begin{table}[!t]
\caption{VitalFR3 FMCW waveform and acquisition timing.}
\label{tab:waveform_configuration}
\centering
\renewcommand{\arraystretch}{1.15}
\begin{tabular}{ll}
\hline
\textbf{Parameter} & \textbf{Value} \\
\hline
Sweep direction
    & Up-chirp \\
Designed sweep bandwidth
    & $100~\mathrm{MHz}$ \\
Designed complex I/Q chirp span
    & $-50$ to $+50~\mathrm{MHz}$ relative to IF \\
Active samples per chirp
    & $65{,}536$ \\
Idle samples per chirp
    & $16{,}384$ \\
Samples per CRI
    & $81{,}920$ \\
Active chirp duration
    & $524.288~\mu\mathrm{s}$ \\
Idle duration
    & $131.072~\mu\mathrm{s}$ \\
CRI
    & $655.36~\mu\mathrm{s}$ \\
Chirps per frame
    & $15$ \\
Samples per frame
    & $1{,}228{,}800$ \\
Frame duration
    & $9.8304~\mathrm{ms}$ \\
Per-position slow-time rate
    & $101.725260~\mathrm{Hz}$ \\
\hline
\end{tabular}
\end{table}

For the nominal waveform design, let $B$ denote the designed sweep
bandwidth and $T_{\mathrm{sw}}$ the active sweep duration. At the X440
complex I/Q interface, one transmitted up-chirp can be represented,
apart from a constant complex amplitude, as
\begin{equation}
s_{\mathrm{IQ}}(t)
=
\exp\!\left\{
j2\pi
\left[
-\frac{B}{2}t
+
\frac{1}{2}S t^2
\right]
\right\},
\qquad
0\leq t<T_{\mathrm{sw}},
\label{eq:tx_fmcw}
\end{equation}
where $t$ is referenced to the start of the active sweep and
$S=B/T_{\mathrm{sw}}$ is the designed FMCW chirp slope.
For $B=100~\mathrm{MHz}$ and $T_{\mathrm{sw}}=524.288~\mu\mathrm{s}$,
the nominal slope is $S=190.735~\mathrm{kHz/\mu s}$.

Because the 15-chirp sample plan is repeated from frame to frame, each
fixed chirp position recurs once every $9.8304~\mathrm{ms}$. Therefore,
the samples associated with a given chirp position form a separate
slow-time sequence with sampling rate
$f_{\mathrm{slow}}=101.725260~\mathrm{Hz}$. The 15 chirp positions are
retained as separate slow-time channels for subsequent target-centric
processing rather than being concatenated into a single sequence at
the chirp repetition rate.

\begin{figure}[t]
    \centering

    \subfloat[]{%
        \includegraphics[width=0.48\columnwidth]
        {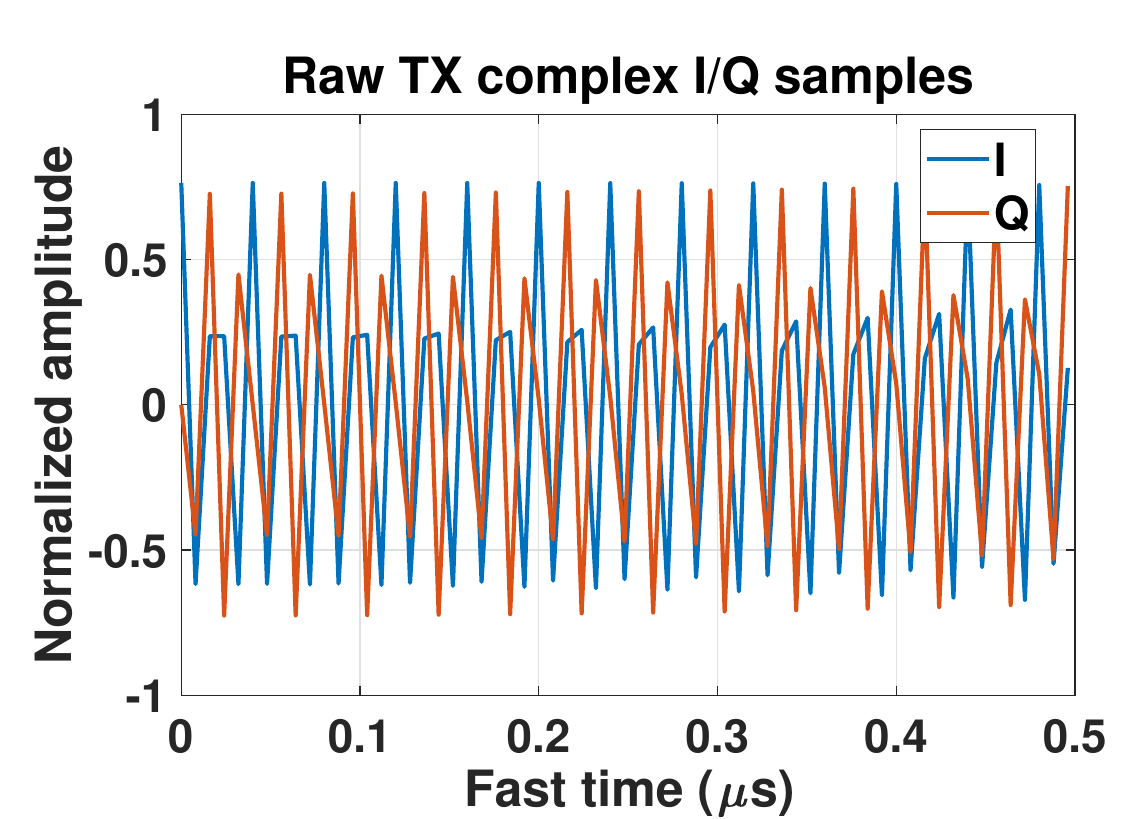}
        \label{fig:tx_waveform_iq}
    }
    \hfill
    \subfloat[]{%
        \includegraphics[width=0.48\columnwidth]
        {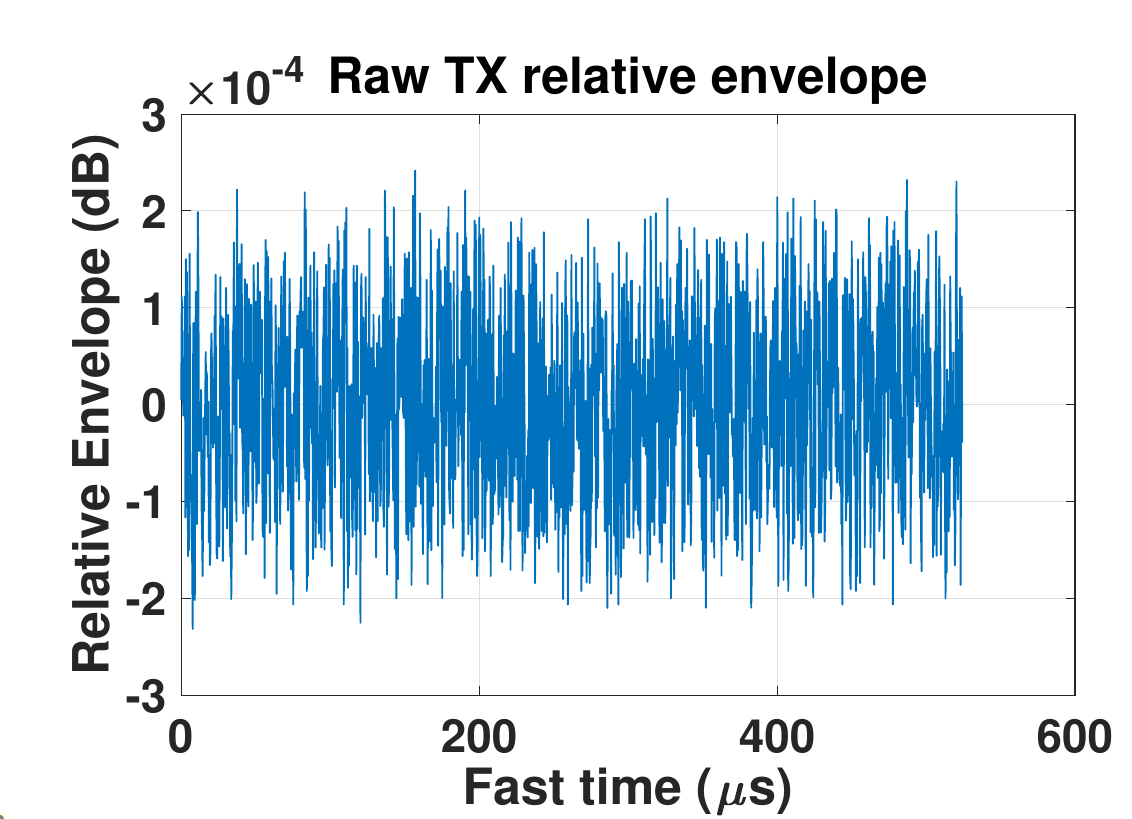}
        \label{fig:tx_waveform_envelope}
    }

    \vspace{0.4em}

    \subfloat[]{%
        \includegraphics[width=0.48\columnwidth]
        {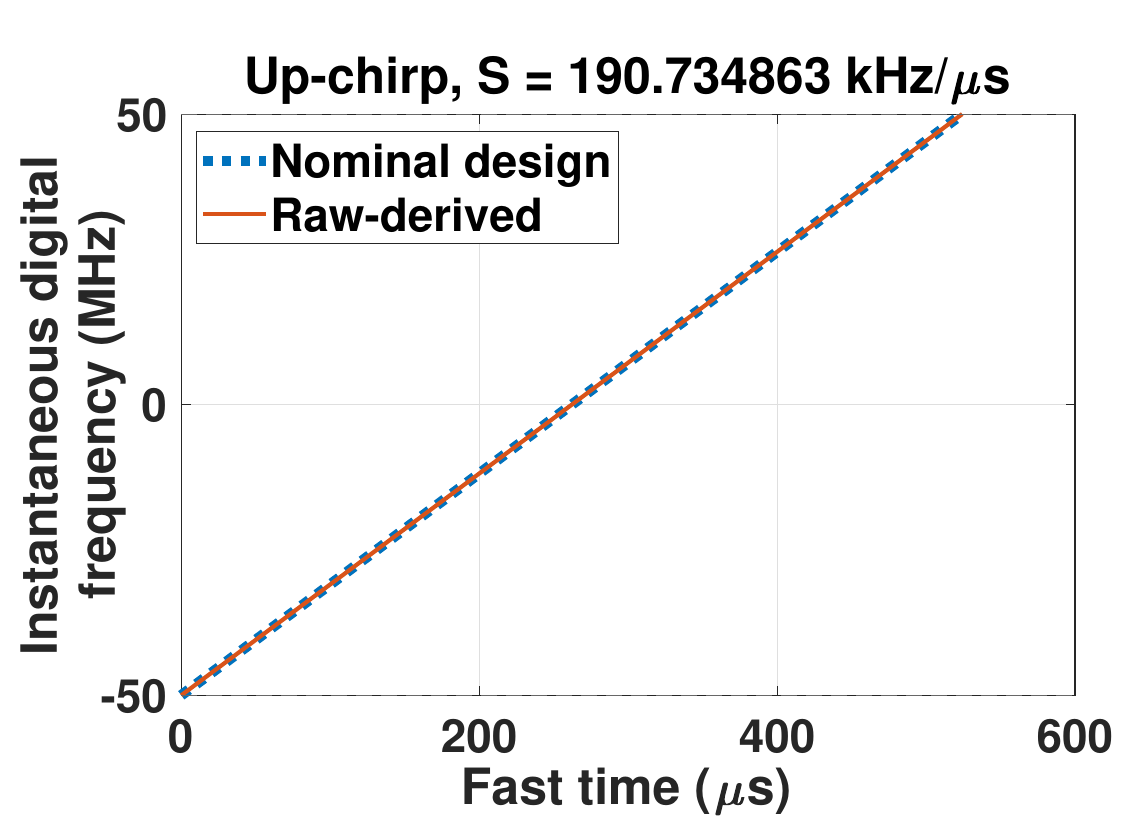}
        \label{fig:tx_waveform_frequency}
    }
    \hfill
    \subfloat[]{%
        \includegraphics[width=0.48\columnwidth]
        {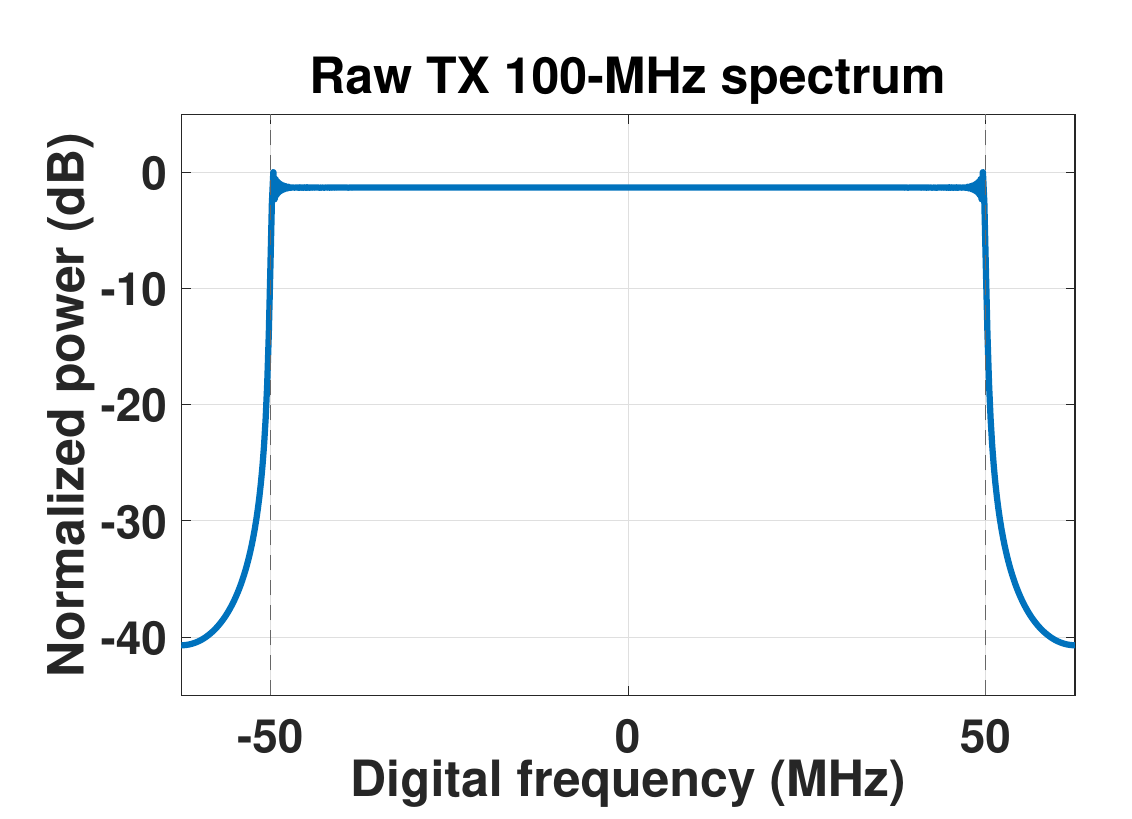}
        \label{fig:tx_waveform_spectrum}
    }

    \caption{Raw transmit waveform verification at the X440 complex I/Q
    interface: (a) recorded complex I/Q samples; (b) transmit envelope
    relative to its median; (c) instantaneous frequency of the up-chirp;
    and (d) normalized digital power spectrum.}
    \label{fig:tx_waveform_validation}
\end{figure}

The realized transmit waveform was verified directly from
the raw complex I/Q transmit (TX) reference at the X440 interface,
independently of the subsequent characterization of the complete
measurement path. The recorded I/Q
samples are shown in Fig.~\ref{fig:tx_waveform_iq}, while
Fig.~\ref{fig:tx_waveform_envelope} shows the corresponding complex
envelope over the active sweep. Its magnitude remains effectively
constant, with deviations confined to approximately
$\pm0.00024~\mathrm{dB}$ relative to the median. 
The instantaneous frequency in Fig.~\ref{fig:tx_waveform_frequency}
closely follows the nominal linear design law over the complete active
sweep. Fig.~\ref{fig:tx_waveform_spectrum} confirms the intended
$100~\mathrm{MHz}$ digital spectral occupancy.

Quantitatively, the instantaneous frequency obtained from the raw data
spans approximately $-50.000000$ to $+49.998474~\mathrm{MHz}$ over the
active samples. The fitted slope is
$190.734863~\mathrm{kHz/\mu s}$, and the frequency span between the first
and last active samples is $99.998474~\mathrm{MHz}$. This agrees with the
designed $100~\mathrm{MHz}$ chirp. The $65{,}536$ active samples cover
$0\leq t<T_{\mathrm{sw}}$, so the nominal span between the first and last
samples is
$B(65{,}535/65{,}536)=99.998474~\mathrm{MHz}$. The small difference from
$100~\mathrm{MHz}$ is therefore due to the discrete sample positions and
does not indicate a reduction in the designed sweep bandwidth.

\subsection{Calibration of the Complete Measurement Path}
\label{subsec:calibration}
Calibration was performed at the instrument level using raw I/Q data
acquired with the antennas removed and the FR3 transmit and receive
paths connected in loopback through a $30~\mathrm{dB}$ attenuator, rather than by
independently de-embedding individual components.
For each of the six calibrated RF center frequencies, the complex transfer
response of the configured measurement path was estimated from the raw RX
acquisition relative to the raw TX reference over the measured chirp
support. These two endpoints at the X440 interface are identified in
Fig.~\ref{fig:FR3_system_schematic}. The installed ZX75BP-750-S+ receive
filter is therefore included in the measured response of the complete
measurement path and is not separately inverted or de-embedded.
The resulting complex response for each RF center provides the basis
for the equalizer derived below.

Before estimating this response, the raw receive (RX) recordings were aligned
and checked directly against the raw TX reference.
An integer lag between RX and TX was first estimated by
normalized complex correlation and then checked across multiple complete
frames and chirp positions to verify that the alignment remained stable.
The digital spectral orientation and dechirp sign were also verified
directly from the raw loopback data rather than assumed from the nominal
frequency plan. A test with a positive delay established
$\operatorname{conj}(\mathrm{RX})\,\mathrm{TX}$ as the dechirp
convention used in the subsequent processing.

For each calibrated RF center frequency, the complex response
was estimated from 15 segments distributed over the loopback recording,
using chirps 1, 8, and 15 from five frame locations at fractions
0.05, 0.25, 0.50, 0.75, and 0.95 of the acquisition. After the
validated integer alignment, the instantaneous complex response of each
segment was formed as
\begin{equation}
H_{\mathrm{inst}}\!\left(f(t)\right)
=
\frac{x_{\mathrm{RX}}(t)}
     {x_{\mathrm{TX}}(t)},
\label{eq:fullchain_response}
\end{equation}
where \(f(t)\) is the instantaneous digital frequency derived from
the raw transmit chirp.
An affine phase component was estimated from the unwrapped response phase
over the central $\pm40~\mathrm{MHz}$ reference region and removed from
the response phase shape. The corresponding residual path delay obtained
from the phase slope was retained for the electrical reference calculation
described in Section~\ref{subsec:calibration_realization}. The magnitude
and residual phase responses were then summarized in 2048 frequency bins
across the measured chirp support and combined across the 15 segments
using the median.

Before constructing the equalizer, the repeatability of the measured response
was evaluated across the 15 distributed segments
at each RF center frequency. For every frequency bin, the standard
deviation of the measured magnitude and residual phase was computed across the 15 segments. Repeatability thresholds were obtained
from the central $\pm40~\mathrm{MHz}$ reference region using the median
plus six median absolute deviation (MAD) terms, with minimum thresholds of
$0.10~\mathrm{dB}$ for magnitude and $1.50^\circ$ for phase. 
A bin was classified as calibratable only if its TX amplitude was no more
than $1.5~\mathrm{dB}$ below the median TX amplitude over the central
$\pm40~\mathrm{MHz}$ reference region. Both repeatability measures also
had to satisfy their respective thresholds.
The response was accepted over the full band only when at least 99.9\% of the measured chirp bins were calibratable and both bins at the chirp edges
satisfied the same criteria.
Up to four non-calibratable bins permitted by this acceptance policy were
interpolated only for equalizer design, while the measured response itself
was retained unchanged.

\begin{figure}[t]
    \centering
    \begin{tabular}{@{}c@{\hspace{0.02\columnwidth}}c@{}}

        \subfloat[]{%
            \includegraphics[width=0.47\columnwidth]
            {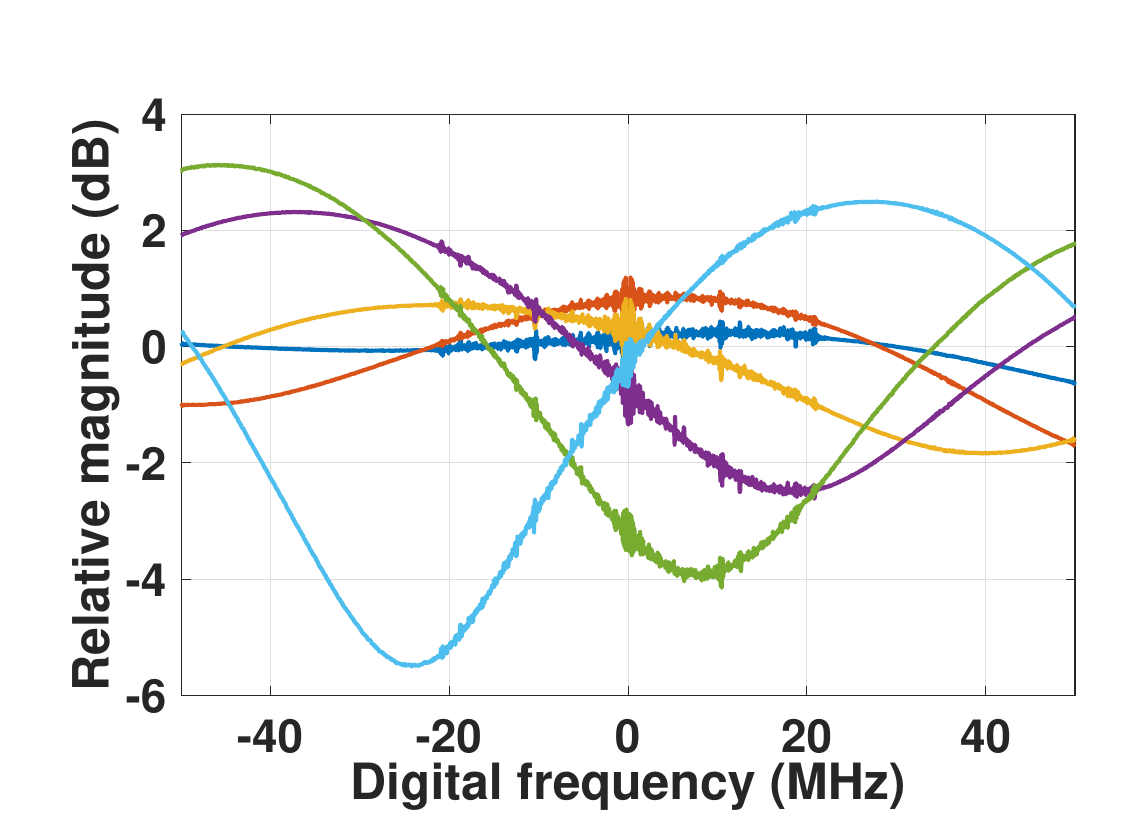}
            \label{fig:cal_measured_mag}
        }
        &
        \subfloat[]{%
            \includegraphics[width=0.47\columnwidth]
            {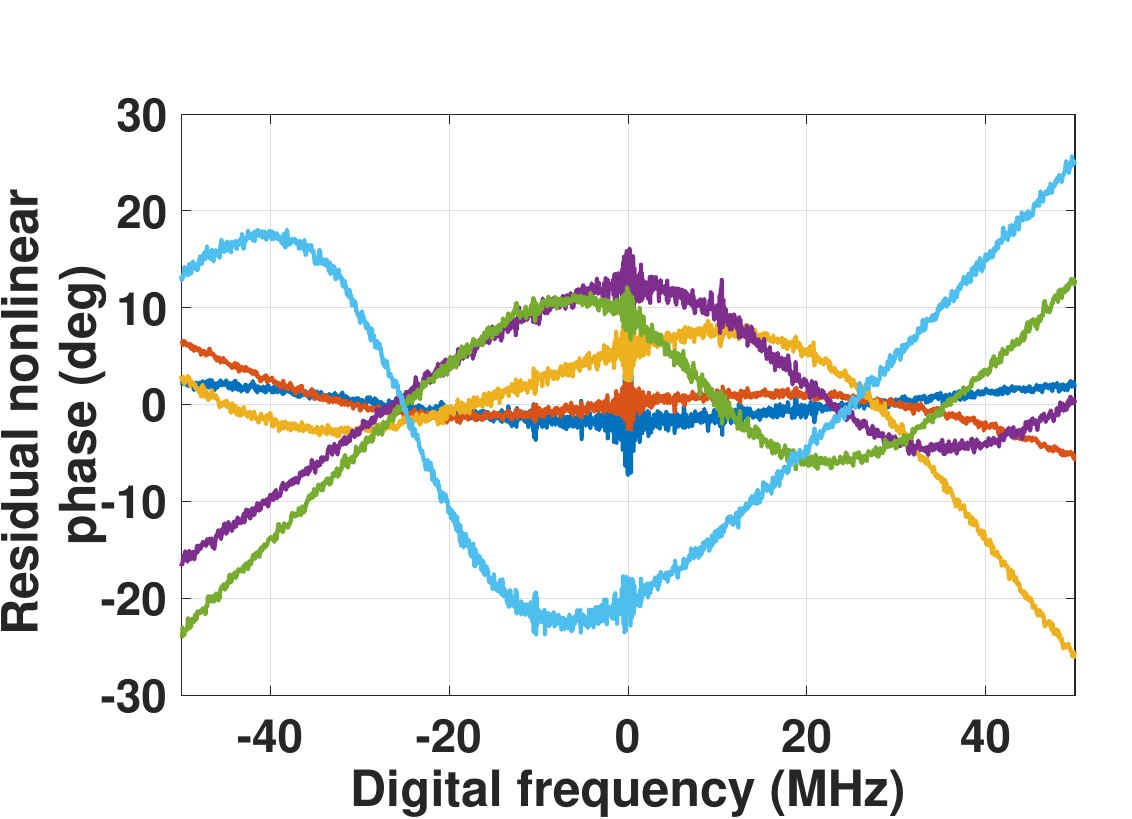}
            \label{fig:cal_measured_phase}
        }
        \\[0.4em]

        \subfloat[]{%
            \includegraphics[width=0.47\columnwidth]
            {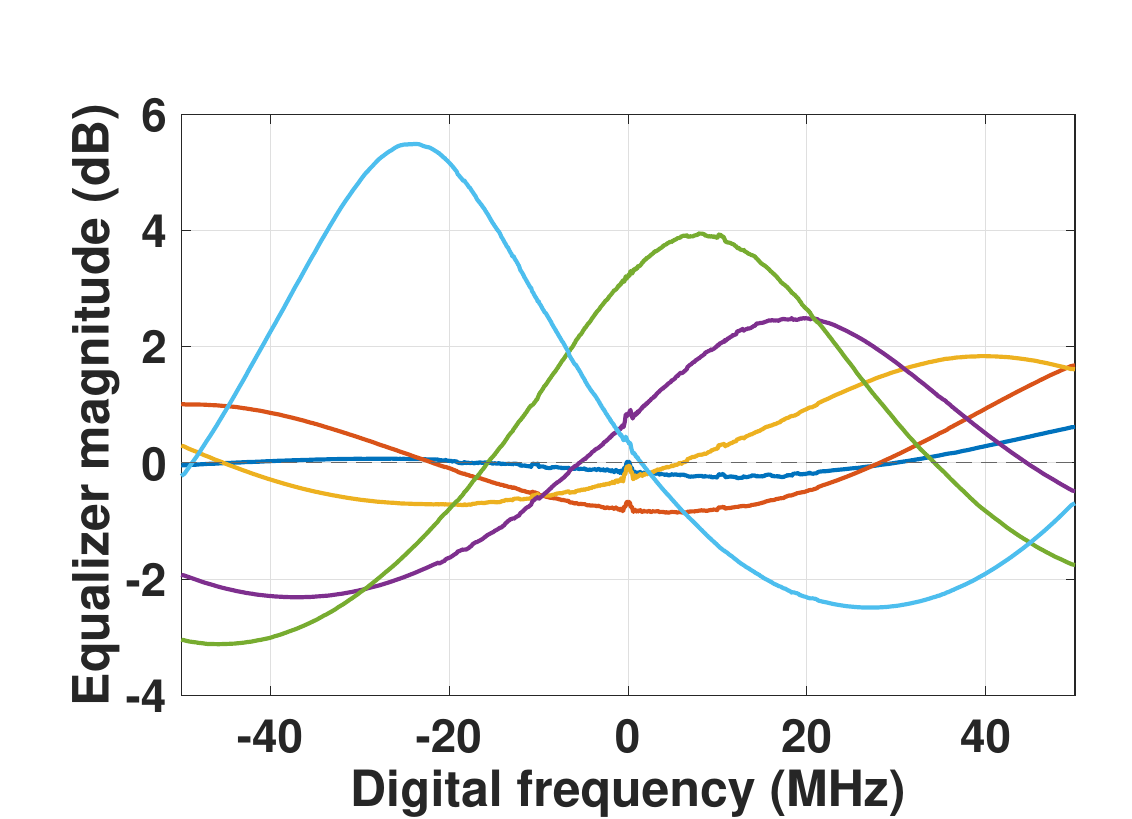}
            \label{fig:cal_eq_mag}
        }
        &
        \subfloat[]{%
            \includegraphics[width=0.47\columnwidth]
            {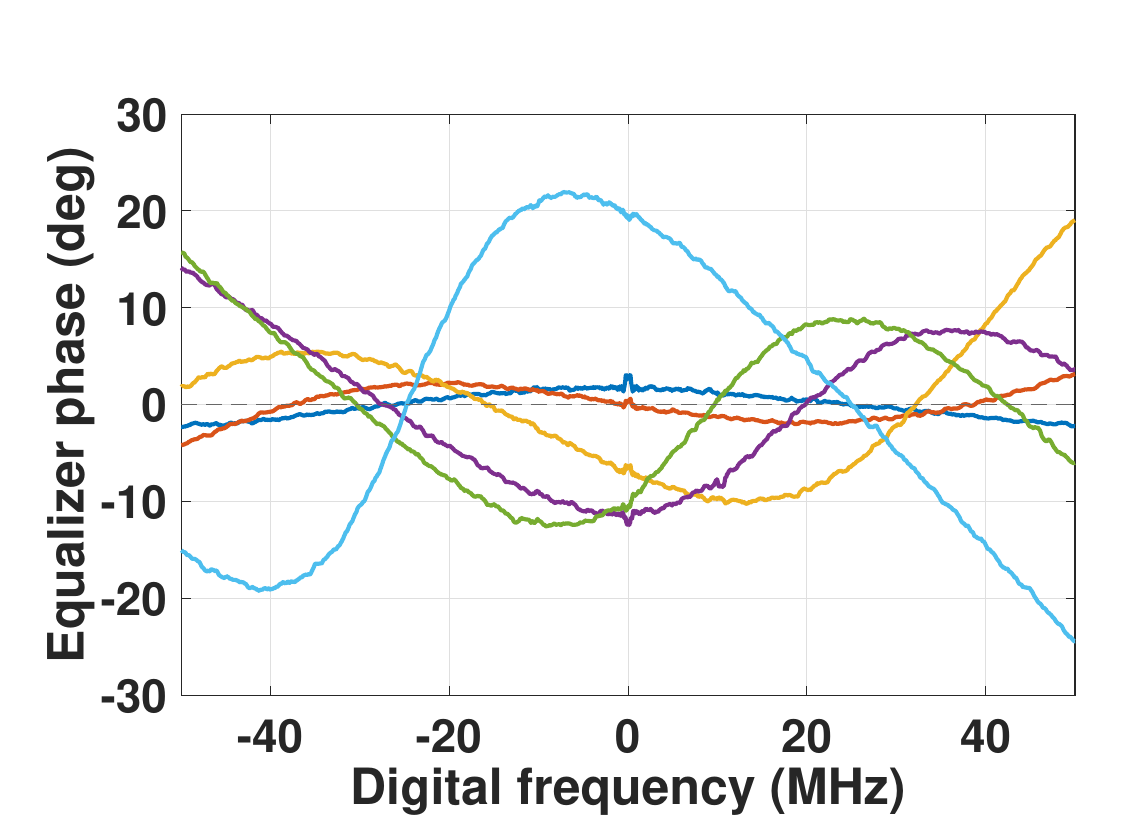}
            \label{fig:cal_eq_phase}
            }
    \end{tabular}
    \par
    \centering
    \includegraphics[width=0.96\columnwidth]{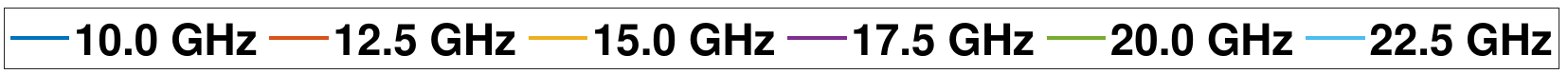}
    \par\smallskip
    \caption{Measured response of the complete measurement path and
    corresponding complex equalizer for each RF center over the measured
    chirp support:
    (a) measured magnitude shape relative to the preserved band gain;
    (b) measured residual nonlinear phase;
    (c) equalizer magnitude correction;
    (d) equalizer phase correction after removal of the affine component.}
    \label{fig:fullchain_equalizer}
\end{figure}

For equalizer design, the measured response was separated into its
RF-dependent bulk gain and its frequency-dependent complex shape. The
median magnitude over the symmetric measured chirp support was retained
as the band gain for each RF center, while the corresponding normalized magnitude
and residual phase shapes were used for correction. A constant phase
term was likewise removed from the residual phase response. 
The resulting response shapes derived from the raw data exhibit
RF-dependent magnitude and nonlinear phase structure across the measured
chirp support, as shown in Figs.~\ref{fig:cal_measured_mag} and
\ref{fig:cal_measured_phase}.
The response shapes used for equalizer construction were smoothed over a
$0.75~\mathrm{MHz}$ span to form the complex response model
$H_{\mathrm{m}}(f)$.

A regularized complex inverse was constructed as
\begin{equation}
E_{\lambda}(f)
=
\frac{H_{\mathrm{m}}^{*}(f)}
     {|H_{\mathrm{m}}(f)|^{2}+\lambda},
\label{eq:regularized_equalizer}
\end{equation}
where $\lambda$ is the regularization parameter controlling the tradeoff
between response flattening and excessive inverse gain in frequency
regions where the measured response is weak. It was selected from
17 logarithmically spaced values over $10^{-8}$--$10^{0}$. For each candidate, 
the equalizer was normalized so that the median
magnitude of the combined response over the symmetric measured chirp
support was unity. 
Candidates whose maximum boost exceeded $6~\mathrm{dB}$
were rejected. Among the admissible candidates, the value of $\lambda$ minimizing the
residual root-mean-square (RMS) magnitude error on the measured response
shape was selected.
Finally, the affine phase component of the selected equalizer was
removed so that the equalizer corrected the nonlinear complex response
shape without introducing an artificial bulk delay. The resulting equalizer magnitude and phase corrections for each RF center after removal of the affine component are shown
in Figs.~\ref{fig:cal_eq_mag} and \ref{fig:cal_eq_phase},
respectively.

\subsection{Calibration Realization, Timing, and Electrical Reference}
\label{subsec:calibration_realization}
For each RF center, the frequency-domain equalizer was implemented as a
$257$-tap causal complex finite impulse response (FIR) filter. The equalizer
correction was applied over the full measured chirp support and was smoothly
transitioned to unity outside the chirp band, reaching unity by
$|f|=60~\mathrm{MHz}$. The FIR introduces a fixed delay of
$D_{\mathrm{FIR}}=128$ samples, corresponding to $1.024~\mu\mathrm{s}$
at $125~\mathrm{MS/s}$, which is compensated after filtering. The filters
are applied continuously, with their state preserved across processing blocks
to avoid boundary transients.

The coarse RX--TX sample alignment established from the raw loopback
measurements is represented by the integer delay $D_{\mathrm{HW}}$ for each RF center.
It accounts for the fixed sample offset between the
transmit and receive paths and is distinct from the $128$-sample delay
$D_{\mathrm{FIR}}$ introduced by the causal calibration FIR. The two
delays are therefore compensated separately.

After the integer RX--TX alignment, a small residual linear phase slope
remained in the loopback response. For each RF center, the slope was
estimated from the unwrapped loopback phase $\phi(f)$ over the central
$\pm40~\mathrm{MHz}$ reference region and converted to a fractional
electrical delay
\begin{equation}
\tau_0
=
-\frac{1}{2\pi}\frac{d\phi}{df}.
\label{eq:residual_electrical_delay}
\end{equation}
The reported value of $\tau_0$ was taken as the median over the same 15
distributed loopback segments used to characterize the complex response.
Using the chirp slope $S$ derived from the raw transmit data, the
corresponding electrical beat-frequency offset for each RF center was
defined as
\begin{equation}
f_0 = S\tau_0 .
\label{eq:electrical_beat_offset}
\end{equation}
The resulting values of $D_{\mathrm{HW}}$, $\tau_0$, and $f_0$ for the
six calibrated RF centers are summarized in
Table~\ref{tab:cal_timing_reference}. At $15~\mathrm{GHz}$, $\tau_0$
changes sign because $D_{\mathrm{HW}}$ changes by one sample, leaving the
residual fractional delay with the opposite sign. This sign change does
not indicate a negative physical delay or noncausal behavior.

\begin{table}[t]
\caption{Timing and electrical reference parameters at each RF center.}
\label{tab:cal_timing_reference}
\centering
\renewcommand{\arraystretch}{1.12}
\begin{tabular}{cccc}
\hline
\textbf{RF center} &
$\mathbf{D_{\mathrm{HW}}}$ &
$\boldsymbol{\tau_0}$ &
$\mathbf{f_0}$ \\
\textbf{(GHz)} &
\textbf{(samples)} &
\textbf{(ns)} &
\textbf{(Hz)} \\
\hline
10   & 404 & $-3.331238$ & $-635.383174$ \\
12.5 & 404 & $-3.714944$ & $-708.569410$ \\
15   & 403 & $+3.743185$ & $+713.955946$ \\
17.5 & 404 & $-2.742069$ & $-523.008190$ \\
20   & 404 & $-2.102393$ & $-400.999729$ \\
22.5 & 404 & $-2.741170$ & $-522.836606$ \\
\hline
\end{tabular}
\end{table}

\begin{figure}[t]
    \centering
    \subfloat[]{%
        \includegraphics[width=0.47\columnwidth]
        {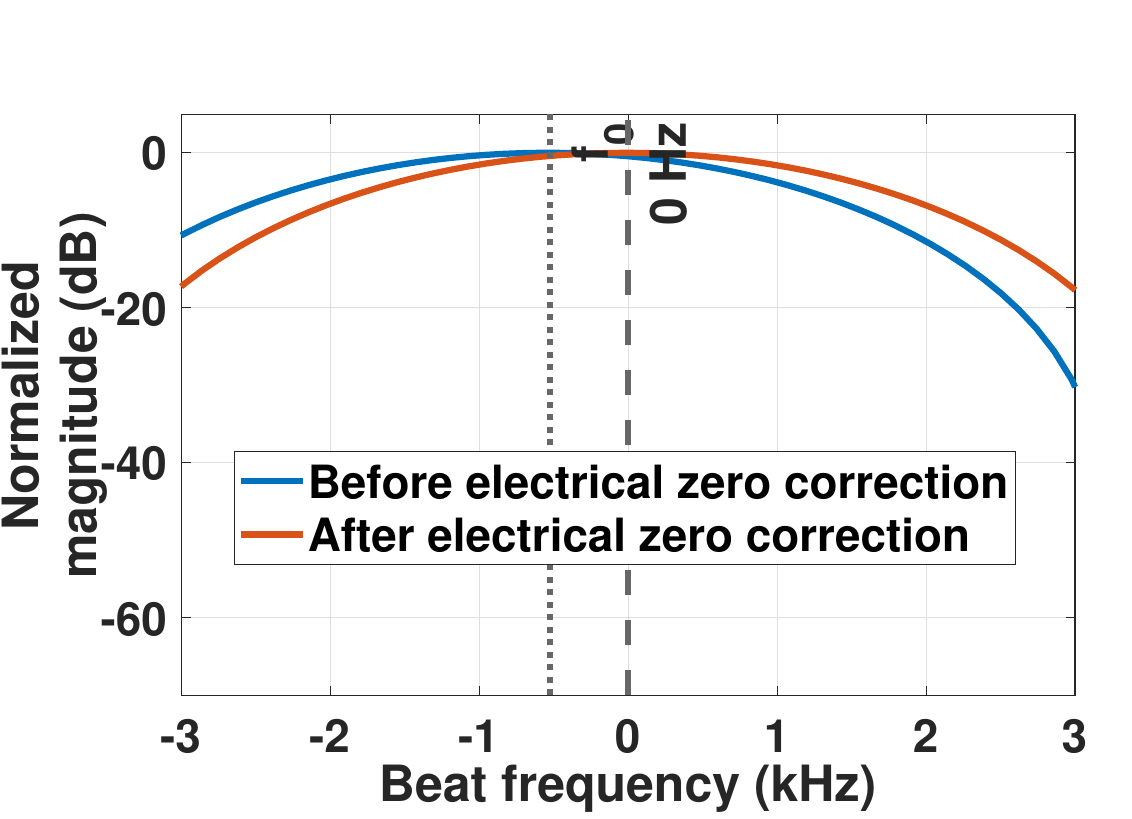}
        \label{fig:FIG_D1a_electrical_zero_beat_22p5GHz}
    }
    \hfill
    \subfloat[]{%
        \includegraphics[width=0.47\columnwidth]
        {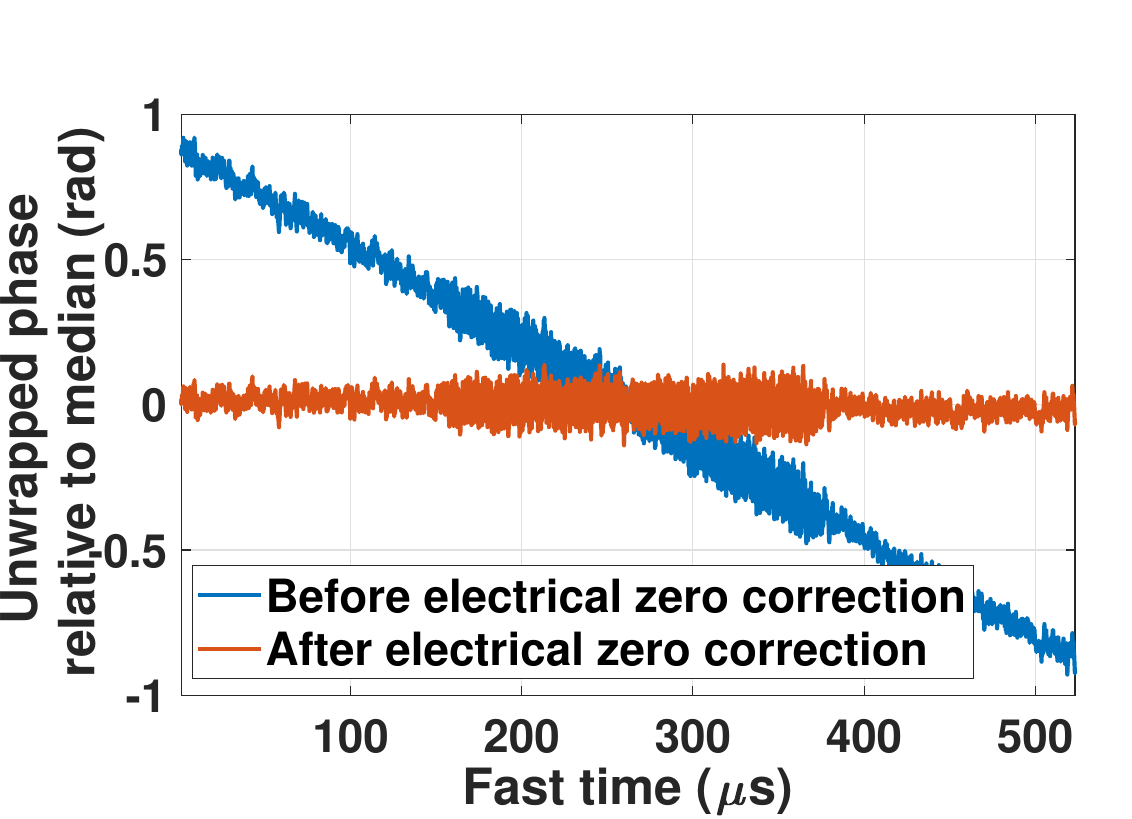}
        \label{fig:FIG_D1b_electrical_zero_phase_22p5GHz}
    }

    \caption{Representative electrical zero correction at
    $22.5~\mathrm{GHz}$: (a) beat frequency before and after correction;
    (b) corresponding phase slope before and after correction.}
    \label{fig:electrical_zero_example}
\end{figure}

The offset $f_0$ for each RF center is applied after dechirping to establish
the electrical beat-frequency zero before subsequent fast-time processing.
This calibration defines only an electrical loopback zero reference; it does
not establish an absolute physical range origin at the antenna aperture.
Its effect is illustrated for a representative $22.5~\mathrm{GHz}$ loopback
measurement in Fig.~\ref{fig:electrical_zero_example}, where the correction
removes the residual beat-frequency offset and the corresponding linear
phase slope.

After dechirping and application of the electrical zero correction for each
RF center, the complex fast-time signal was passed through a four-stage
anti-aliasing (AA) and decimation chain. The successive decimation factors
were $4$, $4$, $4$, and $2$, giving an overall decimation factor of $128$
and reducing the sampling rate from $125~\mathrm{MS/s}$ to
$976.5625~\mathrm{kS/s}$. This produces $512$ complex fast-time samples
over each active chirp while retaining the validated beat-frequency band
$|f_b|\leq300~\mathrm{kHz}$, where $f_b$ denotes the beat frequency after
the electrical zero correction. The delay introduced by each AA stage was
removed before the next stage, leaving no residual delay from the AA filters
in the final output. This AA and decimation chain is distinct from the
$257$-tap calibration FIR described above.

\subsection{Calibration Performance}
\label{subsec:calibration_validation}
The final calibration was assessed in two ways. Across the same 15
distributed loopback segments used for calibration characterization,
all six RF centers had at least $99.95117\%$ calibratable
frequency bins. The RMS error in magnitude shape after correction
remained below $0.061~\mathrm{dB}$ across the calibrated band, compared
with uncorrected values reaching $2.836~\mathrm{dB}$. The finite
$257$-tap FIR realization was also compared with the corresponding
frequency-domain equalizer target. Across the six RF centers, the
worst-case errors were $0.00466~\mathrm{dB}$ in magnitude and
$0.04621^\circ$ in phase. The results for each RF center are summarized
in Table~\ref{tab:calibration_validation}.

\begin{table*}[t]
\caption{Calibration performance and FIR realization accuracy at the
six RF center frequencies.}
\label{tab:calibration_validation}
\centering
\renewcommand{\arraystretch}{1.12}
\begin{tabular}{c c c c c c c}
\hline
\textbf{RF center} &
\textbf{Calibratable} &
\textbf{$\lambda$ (in Eq.~\ref{eq:regularized_equalizer})} &
\textbf{Maximum equalizer} &
\multicolumn{2}{c}{\textbf{RMS error in magnitude shape (dB)}} &
\textbf{FIR realization RMS} \\
\cline{5-6}
\textbf{(GHz)} &
\textbf{fraction (\%)} &
&
\textbf{boost (dB)} &
\textbf{Before} &
\textbf{After} &
\textbf{Magnitude (dB) / Phase ($^\circ$)} \\
\hline
10.0 & 100.00000 & $1.0\times10^{-8}$ & 0.6162 &
0.20088 & 0.06048 & 0.00394 / 0.04621 \\
12.5 & 100.00000 & $3.1623\times10^{-7}$ & 1.6764 &
0.73662 & 0.04073 & 0.00280 / 0.02794 \\
15.0 & 99.95117 & $1.0\times10^{-8}$ & 1.8359 &
0.98115 & 0.06033 & 0.00431 / 0.03923 \\
17.5 & 99.95117 & $1.0\times10^{-8}$ & 2.4908 &
1.70931 & 0.05902 & 0.00466 / 0.04047 \\
20.0 & 99.95117 & $1.0\times10^{-8}$ & 3.9476 &
2.36207 & 0.04763 & 0.00324 / 0.04166 \\
22.5 & 100.00000 & $1.0\times10^{-8}$ & 5.4854 &
2.83575 & 0.04115 & 0.00326 / 0.04108 \\
\hline
\end{tabular}
\end{table*}

\begin{figure}[t]
\centering

\subfloat[]{%
\includegraphics[width=0.49\columnwidth]
{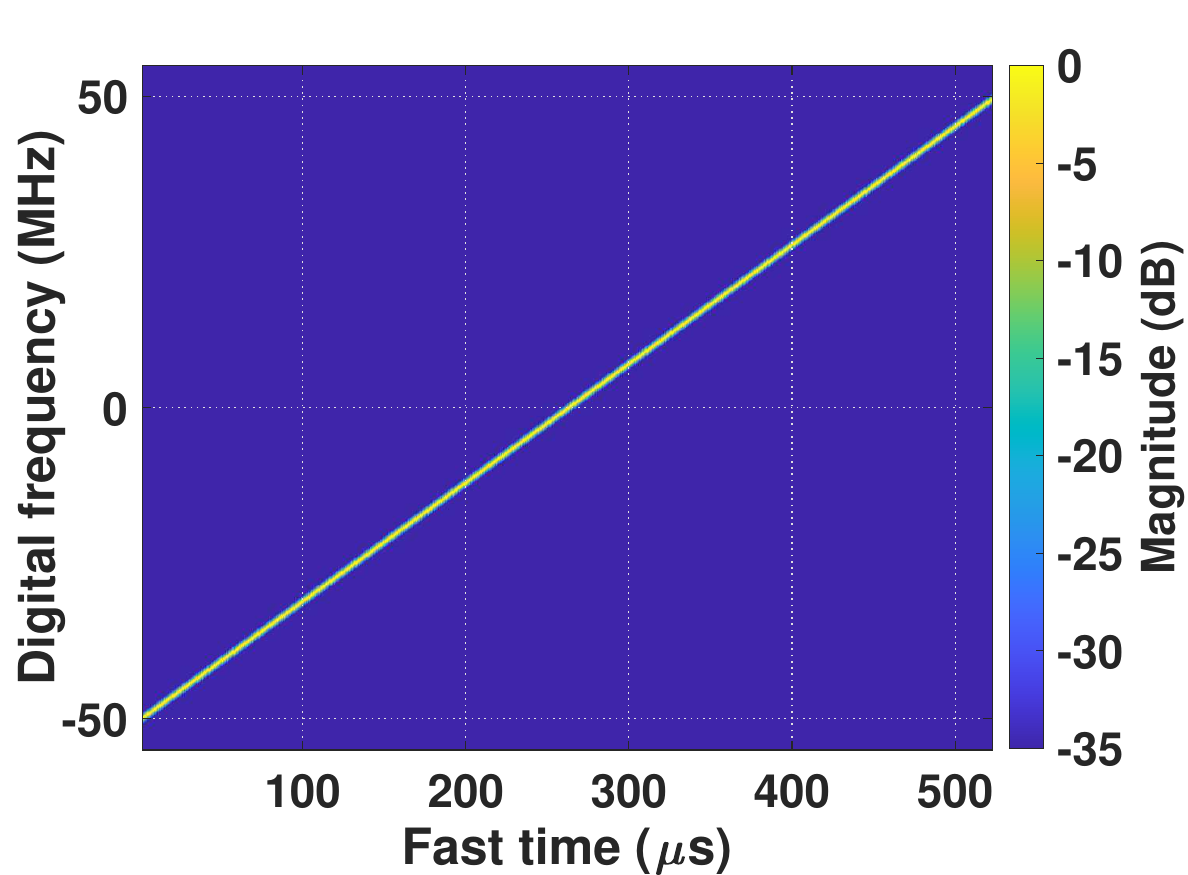}%
\label{fig:tf_tx}}
\hfill
\subfloat[]{%
\includegraphics[width=0.49\columnwidth]
{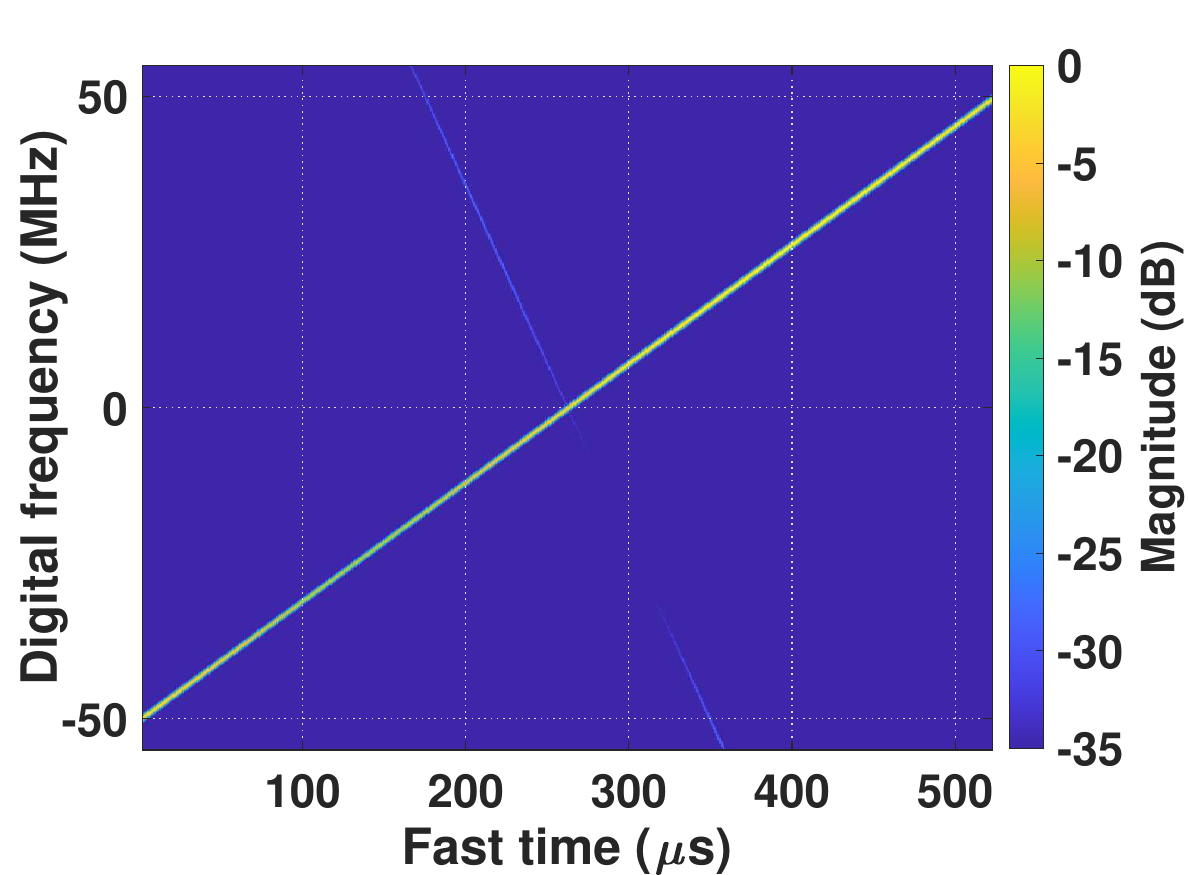}%
\label{fig:tf_rx_raw}}

\vspace{1mm}

\subfloat[]{%
\includegraphics[width=0.49\columnwidth]
{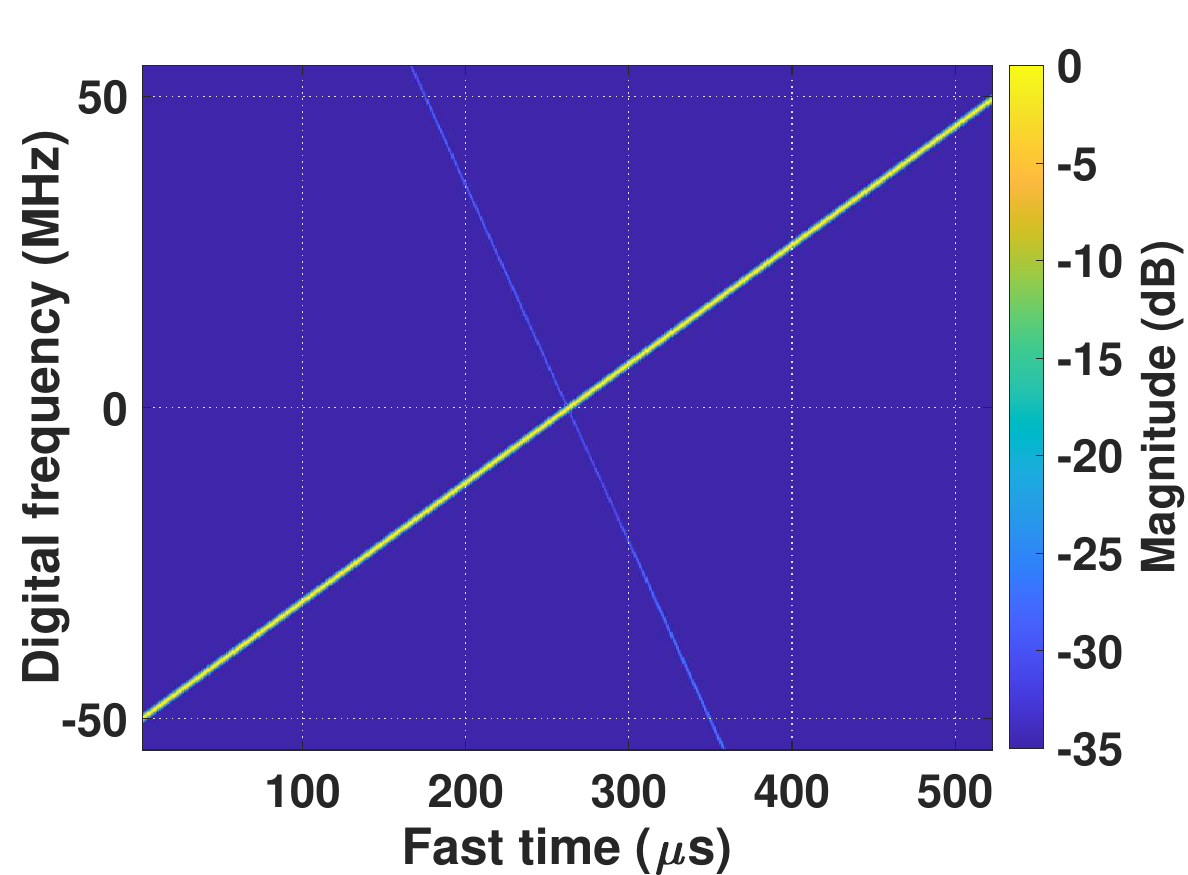}%
\label{fig:tf_rx_calibrated}}
\hfill
\subfloat[]{%
\includegraphics[width=0.49\columnwidth]
{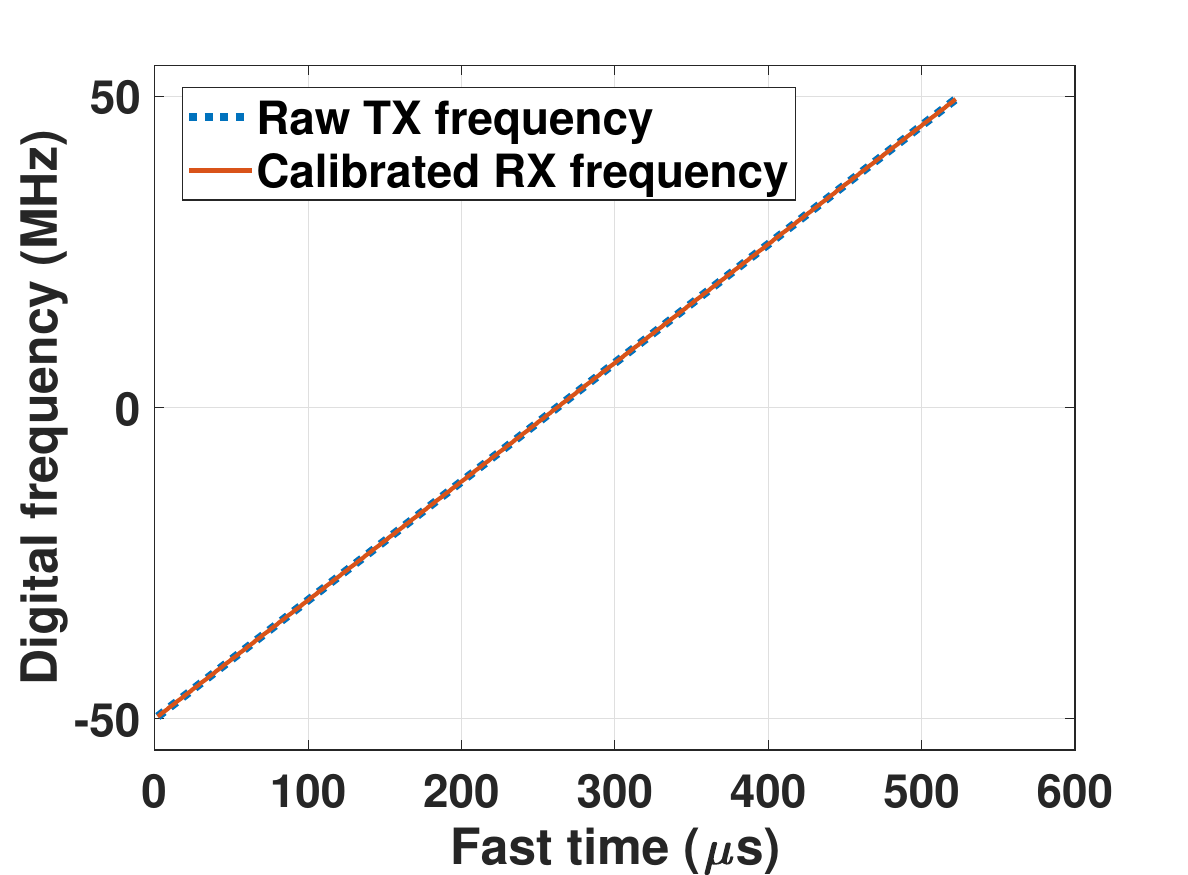}%
\label{fig:tf_chirp_frequency}}

\caption{End-to-end time--frequency validation using a representative
$22.5~\mathrm{GHz}$ loopback measurement:
(a) raw TX spectrogram;
(b) aligned raw RX spectrogram;
(c) RX spectrogram after the final calibration FIR; and
(d) comparison of the raw TX fitted frequency law with the
calibrated RX frequency trajectory.}
\label{fig:tf_validation}
\end{figure}

The FMCW waveform was further evaluated in the time--frequency domain
at all six RF centers. Fig.~\ref{fig:tf_validation} shows the
$22.5~\mathrm{GHz}$ loopback measurement as a representative example.
The calibrated RX signal preserves the positive slope of the raw TX chirp
and closely follows the frequency law fitted to the raw TX data.
Across the six RF centers, the measured
chirp slopes differed from the raw TX reference slope by only a few parts
per million. The anti-aliasing and decimation chain was also validated over
the retained beat-frequency band $|f_b|\leq300~\mathrm{kHz}$, with a
cascade passband ripple of $0.00370~\mathrm{dB}$ and a worst-case
alias-folding level of approximately $-79.14~\mathrm{dB}$.

After calibration and decimation, each acquisition is represented by
calibrated complex fast-time data. These data form the input to the
vital sign processing workflow described in the next section.

\section{Vital Sign Processing Pipeline}
\label{sec:baseline_processing}
\begin{figure*}[!t]
    \centering
    \includegraphics[width=\textwidth]{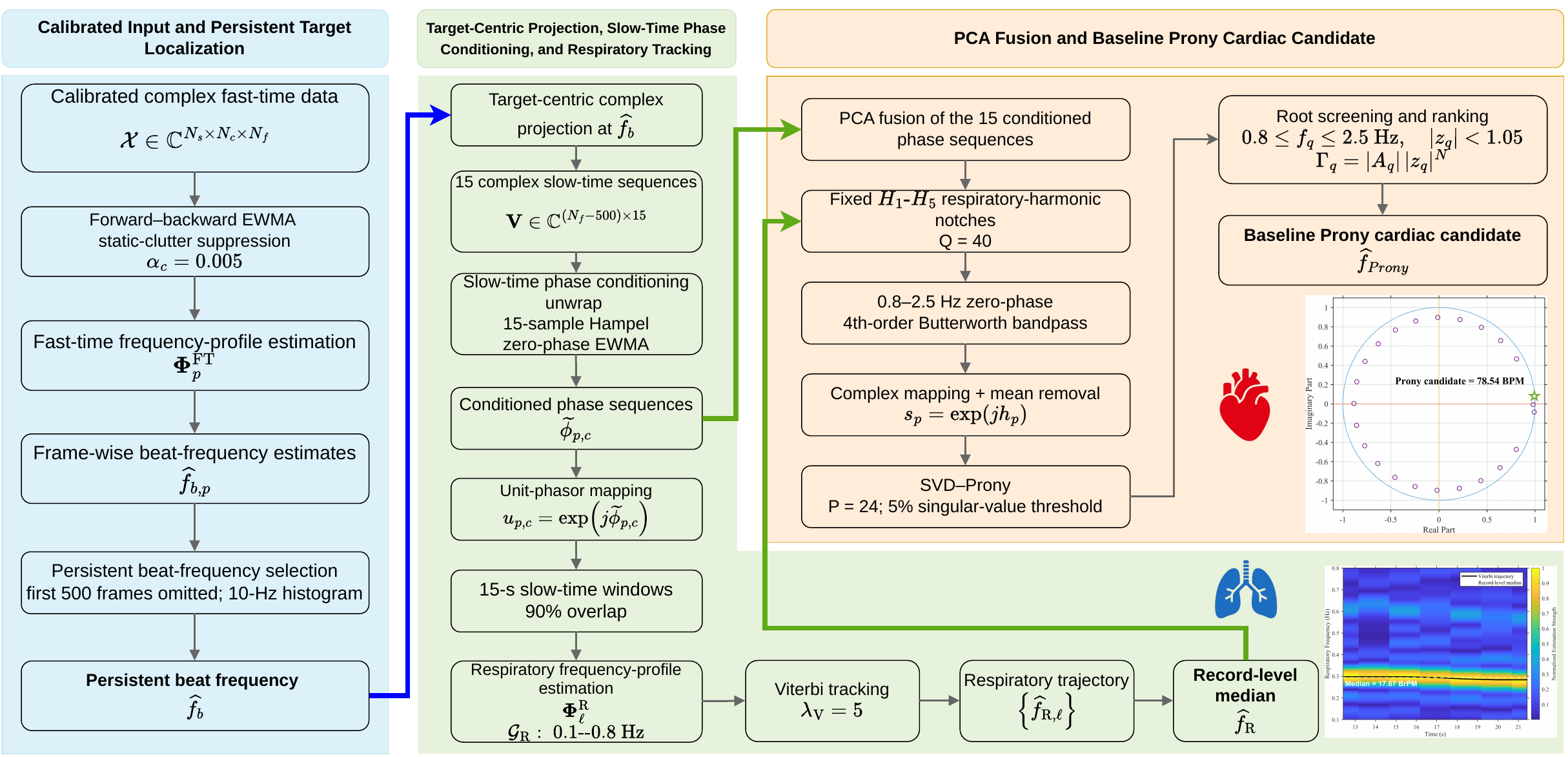}
    \caption{VitalFR3 processing workflow from calibrated
    fast-time data to respiratory tracking and the baseline Prony cardiac
    candidate.}
    \label{fig:baseline_processing}
\end{figure*}

The calibrated complex fast-time data from
Section~\ref{sec:platform} are processed using the workflow shown in
Fig.~\ref{fig:baseline_processing}. The workflow first localizes a
persistent target response in fast-time, then tracks respiration and
estimates the cardiac component from the corresponding slow-time
measurements.

\subsection{Calibrated Fast-Time Input and Static Clutter Suppression}
\label{subsec:baseline_fasttime}
For each acquisition, the calibrated complex fast-time samples are
organized as
\begin{equation}
\mathcal{X}
=
\{x_{m,c,p}\}
\in
\mathbb{C}^{N_s\times N_c\times N_f},
\qquad
N_s=512,\quad N_c=15,
\label{eq:baseline_data_cube}
\end{equation}
where $m$, $c$, and $p$ denote the fast-time sample, chirp position,
and frame indices, respectively, and $N_f$ is the number of acquired
frames. The chirp positions are retained as separate observations
through the subsequent processing stages.

Static clutter is suppressed along the frame dimension using a
forward--backward exponentially weighted moving average (EWMA) with
$\alpha_c=0.005$. Applying the EWMA in both temporal directions avoids introducing a
systematic phase delay into the slow-time sequences.
The resulting data cube after clutter suppression,
$\widetilde{\mathcal{X}}\in\mathbb{C}^{N_s\times N_c\times N_f}$,
is used as the input to the fast-time persistent target response
localization described below.

\subsection{Adopted Frequency-Profile Estimator}
\label{subsec:frequency_profile_estimator}
VitalFR3 uses frequency estimation at two stages of the processing
workflow: fast-time localization of the persistent target response and
slow-time tracking of respiration. For both tasks, we adopt our previously
developed frequency-profile estimator~\cite{11432470}. The
estimator operates directly on the observation snapshots and does not
require explicit covariance estimation or eigen decomposition. Its dominant
computational cost scales linearly with the number of frequency grid
points, snapshots, and temporal samples. These properties make the estimator suitable for repeated use in both
processing stages, while its behavior with nonstationary respiratory
signals is particularly relevant to respiratory tracking. 
For input snapshots $\mathbf{Y}$ and a prescribed frequency grid
$\mathcal{G}=\{f^{(0)},\ldots,f^{(K-1)}\}$, we denote the estimator
output by
\begin{equation}
\boldsymbol{\Phi}
=
\mathcal{E}_{\mathrm{FP}}
\left(\mathbf{Y};\mathcal{G},\mu\right),
\label{eq:frequency_profile_interface}
\end{equation}
where
$\boldsymbol{\Phi}=[\Phi_0,\ldots,\Phi_{K-1}]^{T}$
is the resulting frequency profile and $\mu$ is the estimator step
size. The frequency estimate is obtained from the maximum of the profile as
\begin{equation}
\widehat{f}
=
f^{(k^\star)},
\qquad
k^\star
=
\arg\max_{k\in\{0,\ldots,K-1\}} \Phi_k .
\label{eq:frequency_profile_estimate}
\end{equation}

Further details of the estimator formulation and implementation are
provided in~\cite{11432470}; here, we use the estimator through the
frequency-profile interface defined above.

\subsection{Persistent Beat-Frequency Localization and Target-Centric Projection}
\label{subsec:target_localization}
For fast-time localization, the $N_c=15$ clutter-suppressed chirps in
each frame $p$ are treated as separate fast-time snapshots. The snapshot
corresponding to chirp position $c$ is written as
\begin{equation}
\widetilde{\mathbf{x}}_{c,p}
=
\left[
\widetilde{x}_{0,c,p},
\widetilde{x}_{1,c,p},
\ldots,
\widetilde{x}_{N_s-1,c,p}
\right]^T
\in\mathbb{C}^{N_s},
\label{eq:fasttime_snapshot}
\end{equation}
and the set of snapshots for frame $p$ is
\begin{equation}
\mathbf{Y}^{\mathrm{FT}}_p
=
\left\{
\widetilde{\mathbf{x}}_{c,p}
\right\}_{c=1}^{N_c}.
\label{eq:fasttime_snapshot_set}
\end{equation}
The superscript $\mathrm{FT}$ denotes fast-time observations.

The fast-time search grid $\mathcal{G}_{\mathrm{FT}}$ spans
$0.5$--$10~\mathrm{kHz}$ with $K_{\mathrm{FT}}=1000$ candidate
frequencies. Applying the frequency-profile estimator to
$\mathbf{Y}^{\mathrm{FT}}_p$ with $\mu_{\mathrm{FT}}=0.1$ produces the
fast-time profile $\boldsymbol{\Phi}^{\mathrm{FT}}_p$. The corresponding
frame-wise estimate of the target beat frequency,
$\widehat{f}_{b,p}$, is obtained from the profile maximum.

To exclude the initial acquisition transient, the first 500 frames,
corresponding to approximately $4.9~\mathrm{s}$, are omitted when
selecting the persistent target. The retained frame-wise estimates
$\{\widehat{f}_{b,p}\}$ are then grouped using $10~\mathrm{Hz}$
histogram bins, and the center frequency of the most populated bin is
selected as the persistent target beat frequency, denoted by
$\widehat{f}_b$. This aggregation reduces the influence of isolated
frame-wise deviations.

Once the persistent target beat frequency $\widehat{f}_b$ has been
identified, each retained chirp is projected at this frequency to
obtain a target-centric complex measurement. For chirp position $c$ in
frame $p$, the complex projection is
\begin{equation}
v_{p,c}
=
\sum_{m=0}^{N_s-1}
\widetilde{x}_{m,c,p}
\exp\!\left(
-j2\pi \widehat{f}_b m T_s^{d}
\right),
\label{eq:target_complex_projection}
\end{equation}
where $T_s^{d}$ is the fast-time sampling interval after decimation.
The resulting target-centric complex measurements are arranged as the two-dimensional matrix
$\mathbf{V}=[v_{p,c}]\in\mathbb{C}^{(N_f-500)\times N_c}$,
where the rows correspond to the retained frames and the columns
correspond to the $N_c=15$ chirp positions. Each column of
$\mathbf{V}$ therefore forms a separate complex slow-time sequence
associated with a fixed chirp position.

\subsection{Slow-Time Phase Conditioning and Respiratory Tracking}
\label{subsec:slowtime_respiration}
The phase of each column of $\mathbf{V}$ is independently unwrapped
along slow-time and processed with a 15-sample Hampel filter to suppress
impulsive deviations. Slow baseline variations are then removed using
a zero-phase EWMA with $\alpha_{\phi}=0.00215513$, selected to limit attenuation
at the lower edge of the respiratory band ($0.1~\mathrm{Hz}$) to
approximately $1~\mathrm{dB}$. The forward--backward implementation
avoids introducing a systematic phase shift into the conditioned
slow-time sequences. The resulting phase samples are denoted by
$\widetilde{\phi}_{p,c}$.

For respiratory frequency estimation, the conditioned phase samples are
mapped to unit-modulus complex phasors,
\begin{equation}
u_{p,c}
=
\exp\!\left(
j\widetilde{\phi}_{p,c}
\right).
\label{eq:slowtime_unit_phasor}
\end{equation}
The resulting $N_c=15$ slow-time sequences are retained separately and
used as the observations for respiratory frequency tracking.

A sliding window of $15~\mathrm{s}$ with $90\%$ overlap is applied to
the slow-time sequences in \eqref{eq:slowtime_unit_phasor}. Let $W$
denote the number of slow-time samples in each window and let $p_\ell$
denote the starting frame of the $\ell$th window. For chirp position
$c$, the corresponding slow-time snapshot is
\begin{equation}
\mathbf{u}_{c,\ell}
=
\left[
u_{p_\ell,c},
u_{p_\ell+1,c},
\ldots,
u_{p_\ell+W-1,c}
\right]^T
\in\mathbb{C}^{W},
\label{eq:slowtime_snapshot}
\end{equation}
and the set of observations for that window is
\begin{equation}
\mathbf{Y}^{\mathrm{ST}}_\ell
=
\left\{
\mathbf{u}_{c,\ell}
\right\}_{c=1}^{N_c}.
\label{eq:slowtime_snapshot_set}
\end{equation}
Here, the superscript $\mathrm{ST}$ denotes slow-time observations.

The respiratory search grid
$\mathcal{G}_{\mathrm{R}}
=\{f_{\mathrm{R}}^{(k)}\}_{k=0}^{K_{\mathrm{R}}-1}$
spans $0.1$--$0.8~\mathrm{Hz}$ with
$K_{\mathrm{R}}=800$ candidate frequencies.
For each slow-time window
$\ell$, the frequency-profile estimator is applied to
$\mathbf{Y}^{\mathrm{ST}}_\ell$ with $\mu_{\mathrm{R}}=0.1$ to obtain
the respiratory frequency profile
$\boldsymbol{\Phi}^{\mathrm{R}}_\ell$.

Before Viterbi tracking, the complete respiratory frequency-profile map
is normalized by its global maximum.
Let $\overline{\Phi}^{\mathrm{R}}_{\ell,k}$ denote the normalized
profile value at candidate frequency $f_{\mathrm{R}}^{(k)}$ in the
$\ell$th window, and let $J_{\ell,k}$ denote the cumulative path score.
The Viterbi recursion is
\begin{equation}
\begin{aligned}
J_{1,k}
&=
\overline{\Phi}^{\mathrm{R}}_{1,k},
\\
J_{\ell,k}
&=
\overline{\Phi}^{\mathrm{R}}_{\ell,k}
\\
&\quad+
\max_{0\leq k'<K_{\mathrm{R}}}
\Bigl[
J_{\ell-1,k'}
\\
&\qquad\qquad
-
\lambda_{\mathrm{V}}
\left|
f_{\mathrm{R}}^{(k)}
-
f_{\mathrm{R}}^{(k')}
\right|
\Bigr].
\end{aligned}
\label{eq:viterbi_recursion}
\end{equation}
The recursion is applied for $\ell\geq2$, and
$\lambda_{\mathrm{V}}=5~\mathrm{Hz}^{-1}$ is the
transition penalty weight.
Backtracking from the maximum terminal score gives the optimal state
sequence $\{\widehat{k}_{\ell}\}$, from which the respiratory
frequency trajectory is
$\widehat{f}_{\mathrm{R},\ell}
=f_{\mathrm{R}}^{(\widehat{k}_{\ell})}$.
A representative
respiratory frequency for the complete acquisition is then obtained as
\begin{equation}
\widehat{f}_{\mathrm{R}}
=
\operatorname{median}_{\ell}
\left\{
\widehat{f}_{\mathrm{R},\ell}
\right\}.
\label{eq:representative_resp_frequency}
\end{equation}

\subsection{Principal Component Fusion, Respiratory Harmonic Suppression, and
Baseline Prony Candidate}
\label{subsec:cardiac_candidate}
The mean is first removed from each of the $N_c=15$ conditioned phase
sequences, and principal component analysis (PCA) is then applied to the
resulting phase matrix. The first principal component is retained as
the fused phase observation. Its polarity is aligned with the sample-wise
mean of the 15 conditioned phase sequences, and its amplitude is
rescaled to match the standard deviation of that mean.

Respiratory harmonics are suppressed using five notch filters centered
at $f_{\mathrm{H},n}=n\widehat{f}_{\mathrm{R}}$, $n=1,\ldots,5$.
Each notch has a quality factor of $Q=40$. The filters are designed
once per acquisition using the representative respiratory frequency
$\widehat{f}_{\mathrm{R}}$ and then applied to the fused phase
observation. After notch filtering, the fused phase
observation is band-limited to $0.8$--$2.5~\mathrm{Hz}$ using a
fourth-order Butterworth filter applied in forward--backward form.
The resulting real-valued sequence is denoted by $h_p$.

For Prony analysis, the filtered phase sequence is first mapped to a
unit-modulus complex sequence,
\begin{equation}
s_p
=
\exp\!\left(jh_p\right),
\label{eq:prony_complex_mapping}
\end{equation}
after which its sample mean is removed to obtain
$\widetilde{s}_p$. A linear prediction model of
order $P=24$ is then fitted according to
\begin{equation}
\widetilde{s}_p
+
\sum_{r=1}^{P} a_r \widetilde{s}_{p-r}
\approx 0,
\label{eq:prony_prediction_model}
\end{equation}
where $p$ indexes the slow-time samples, $r$ is the lag
index, and $a_r$ is the corresponding complex linear prediction
coefficient. The coefficients
$\{a_r\}_{r=1}^{P}$ are obtained from the corresponding least-squares
system using singular value decomposition (SVD) regularization, retaining singular values greater than
$5\%$ of the largest singular value.

The roots of the fitted prediction polynomial,
\begin{equation}
z^P
+
\sum_{r=1}^{P} a_r z^{P-r}
=
0,
\label{eq:prony_characteristic_polynomial}
\end{equation}
define the Prony modes.
For the $q$th nonzero root $z_q$, let
$\operatorname{Arg}(z_q)\in(-\pi,\pi]$ denote its principal phase.
The corresponding modal frequency is
\begin{equation}
f_q
=
\frac{f_{\mathrm{slow}}}{2\pi}
\left|
\operatorname{Arg}(z_q)
\right|,
\label{eq:prony_mode_frequency}
\end{equation}
where $f_{\mathrm{slow}}$ is the slow-time sampling rate.
Only roots satisfying
$0.8\leq f_q\leq2.5~\mathrm{Hz}$ and
$0<|z_q|<1.05$ are retained as candidates in the cardiac band.

The complex modal amplitudes $A_q$ are obtained by least-squares
fitting of the Prony modes to $\widetilde{s}_p$. Each valid
root is then ranked using
\begin{equation}
\Gamma_q
=
|A_q|\,|z_q|^{N},
\label{eq:prony_root_ranking}
\end{equation}
where $N$ is the number of slow-time samples used in the Prony
analysis. The valid root with the largest $\Gamma_q$ defines the
baseline Prony cardiac candidate, whose frequency is denoted by
$\widehat{f}_{\mathrm{Prony}}$.
Algorithm~\ref{alg:baseline_vitalfr3} summarizes the steps described above.

\begin{algorithm}[!t]
\caption{VitalFR3 vital sign processing workflow}
\label{alg:baseline_vitalfr3}
\begin{algorithmic}[1]

\Require Calibrated complex fast-time data
\Ensure $\widehat{f}_{\mathrm{R}}$, $\widehat{f}_{\mathrm{Prony}}$

\State Form $\mathcal{X}$; suppress static clutter
$(\alpha_c=0.005)$ to obtain $\widetilde{\mathcal{X}}$

\For{each frame $p$}
    \State Form $\mathbf{Y}_{p}^{\mathrm{FT}}$
    using Eq.~\ref{eq:fasttime_snapshot_set}
    \State Apply Eqs.~\ref{eq:frequency_profile_interface} and
    \ref{eq:frequency_profile_estimate} with $\mu_{\mathrm{FT}}=0.1$
    $\rightarrow\widehat{f}_{b,p}$
\EndFor

\State Discard the first 500 frames
\State $\widehat{f}_b\gets$ center of the most populated
$10$-Hz histogram bin

\For{each retained $(p,c)$}
    \State Compute $v_{p,c}$ using
    Eq.~\ref{eq:target_complex_projection}
\EndFor
\State $\mathbf{V}\gets[v_{p,c}]$

\For{each chirp position $c$}
    \State Unwrap the phase of $v_{p,c}$; Hampel filter (15 samples)
    \State Zero-phase EWMA $(\alpha_{\phi}=0.00215513)$
    $\rightarrow\widetilde{\phi}_{p,c}$
    \State $u_{p,c}\gets\exp(j\widetilde{\phi}_{p,c})$
\EndFor

\For{each $15$-s slow-time window $\ell$ with $90\%$ overlap}
    \State Form $\mathbf{Y}_{\ell}^{\mathrm{ST}}$
    using Eq.~\ref{eq:slowtime_snapshot_set}
    \State Apply Eq.~\ref{eq:frequency_profile_interface} with
    $\mu_{\mathrm{R}}=0.1$
    $\rightarrow\boldsymbol{\Phi}_{\ell}^{\mathrm{R}}$
\EndFor

\State Obtain $\{\widehat{f}_{\mathrm{R},\ell}\}$ by Viterbi tracking
using Eq.~\ref{eq:viterbi_recursion}
\State Compute $\widehat{f}_{\mathrm{R}}$ using
Eq.~\ref{eq:representative_resp_frequency}

\State Mean-center the $N_c$ conditioned phase sequences and apply PCA
\State Align and rescale the first principal component

\State Apply notch filters at
$f_{\mathrm{H},n}=n\widehat{f}_{\mathrm{R}}$,
$n=1,\ldots,5$, with $Q=40$
\State Apply a fourth-order Butterworth filter over
$0.8$--$2.5~\mathrm{Hz}$ in forward--backward form
$\rightarrow h_p$

\State $s_p\gets\exp(jh_p)$
\State Remove the sample mean from $s_p$ to obtain $\widetilde{s}_p$

\State Fit Eq.~\ref{eq:prony_prediction_model} with $P=24$;
retain singular values $>5\%$ of the largest

\State Compute $z_q$ from
Eq.~\ref{eq:prony_characteristic_polynomial}
\State Compute $f_q$ using
Eq.~\ref{eq:prony_mode_frequency}

\State Estimate $A_q$ by least-squares fitting of the
Prony modes to $\widetilde{s}_p$
\State Retain $0.8\leq f_q\leq2.5~\mathrm{Hz}$,
$0<|z_q|<1.05$

\State Rank valid roots using
Eq.~\ref{eq:prony_root_ranking}
\State $\widehat{f}_{\mathrm{Prony}}\gets$
frequency of the highest-ranked valid root

\end{algorithmic}
\end{algorithm}

\section{Experimental Physiological Validation}
\label{sec:experimental_validation}
The processing workflow described in
Section~\ref{sec:baseline_processing} is evaluated using 16 radar
measurement records. A separate end-expiratory apnea acquisition is
examined as a positive control. We also characterize cases in which
the Prony cardiac candidate approaches respiratory harmonics.

\subsection{Experimental Data, References, and Evaluation Roles}
\label{subsec:experimental_data}
The evaluation comprises 16 radar measurement records, identified as
Records 01--16. All records were acquired from a single participant in
separate acquisitions spanning different days and times of day. Clothing
and environmental conditions varied across the measurements. The record
identifiers are used only to distinguish the measurements and do not
indicate their temporal acquisition order.

The measurements were conducted in an indoor laboratory of approximately
$50~\mathrm{m}^2$, with the participant seated approximately $1~\mathrm{m}$
from the radar, as described in Section~\ref{sec:platform}. Respiratory
reference measurements were acquired using a Vernier Go Direct Respiration
Belt, while the cardiac reference was obtained from electrocardiogram (ECG)
recordings acquired with an Analog Devices MAX-ECG-MONITOR.
For record-level respiratory evaluation, the signal from
the respiration belt was detrended and passed through a low-pass filter.
The dominant spectral component within the respiratory frequency range
served as the reference respiratory frequency. For each record, the cardiac
reference was the median heart rate from the corresponding ECG recording.

The physiological reference signals were not used in the radar
processing described in Section~\ref{sec:baseline_processing}. The radar
and physiological reference sensors did not share a common acquisition
trigger or time base. Therefore, we compare the respiratory and cardiac
rates for each recording rather than individual beats or samples.

A separate end-expiratory apnea acquisition was collected as a positive
control. It is not included among Records 01--16.

\subsection{Baseline Vital Sign Performance}
\label{subsec:baseline_performance}

Table~\ref{tab:baseline_record_results} summarizes the respiratory and
cardiac results for all 16 measurements. The respiratory rate estimates
had an MAE of $0.758~\mathrm{BrPM}$ and a median
absolute error of $0.472~\mathrm{BrPM}$. In 15 of the 16 records,
the absolute respiratory error was no greater than $1.114~\mathrm{BrPM}$.
Record~03 was the exception, with an absolute error of
$4.585~\mathrm{BrPM}$.
For cardiac estimation, the baseline Prony candidate was within
$\pm5~\mathrm{BPM}$ of the ECG reference in 11 of the 16 records
($68.75\%$). Across all records, the cardiac MAE was
$4.650~\mathrm{BPM}$, the median absolute error was
$3.424~\mathrm{BPM}$, and the maximum absolute error was
$15.586~\mathrm{BPM}$.

\begin{table*}[!t]
\centering
\caption{Record-level baseline respiratory and cardiac performance.
Physiological references are used only for post hoc evaluation.}
\label{tab:baseline_record_results}
\small
\setlength{\tabcolsep}{4.5pt}
\begin{tabular}{c c c c c c c c}
\hline
Record &
RF &
\multicolumn{3}{c}{Respiration (BrPM)} &
\multicolumn{3}{c}{Cardiac (BPM)} \\
\cline{3-5}\cline{6-8}
&
(GHz) &
Reference &
Radar &
Abs. error &
ECG reference &
Prony candidate &
Abs. error \\
\hline
01 & 10   & 18.164 & 17.670 & 0.494 & 79.050 & 78.536 & 0.514 \\
02 & 12.5 & 18.677 & 18.563 & 0.114 & 87.000 & 74.918 & 12.082 \\
03 & 17.5 & 18.604 & 23.189 & 4.585 & 85.900 & 86.719 & 0.819 \\
04 & 20   & 18.677 & 18.038 & 0.639 & 86.000 & 92.607 & 6.607 \\
05 & 10   & 18.237 & 19.194 & 0.957 & 91.200 & 80.202 & 10.998 \\
06 & 12.5 & 18.823 & 18.195 & 0.628 & 90.100 & 83.625 & 6.475 \\
07 & 15   & 18.750 & 18.300 & 0.450 & 90.400 & 74.814 & 15.586 \\
08 & 17.5 & 18.530 & 18.143 & 0.388 & 89.900 & 94.589 & 4.689 \\
09 & 20   & 18.237 & 19.352 & 1.114 & 88.900 & 87.614 & 1.286 \\
10 & 22.5 & 18.970 & 18.458 & 0.512 & 88.750 & 89.726 & 0.976 \\
11 & 10   & 21.753 & 21.612 & 0.141 & 93.800 & 89.038 & 4.762 \\
12 & 12.5 & 20.654 & 19.667 & 0.987 & 93.600 & 94.292 & 0.692 \\
13 & 15   & 19.189 & 19.089 & 0.101 & 92.700 & 94.705 & 2.005 \\
14 & 17.5 & 19.702 & 19.299 & 0.403 & 91.800 & 91.862 & 0.062 \\
15 & 20   & 19.336 & 19.615 & 0.279 & 91.000 & 87.844 & 3.156 \\
16 & 22.5 & 19.482 & 19.825 & 0.342 & 90.800 & 87.108 & 3.692 \\
\hline
\end{tabular}
\end{table*}

\begin{figure}[!t]
\centering

\subfloat[]{
\includegraphics[width=0.47\columnwidth]
{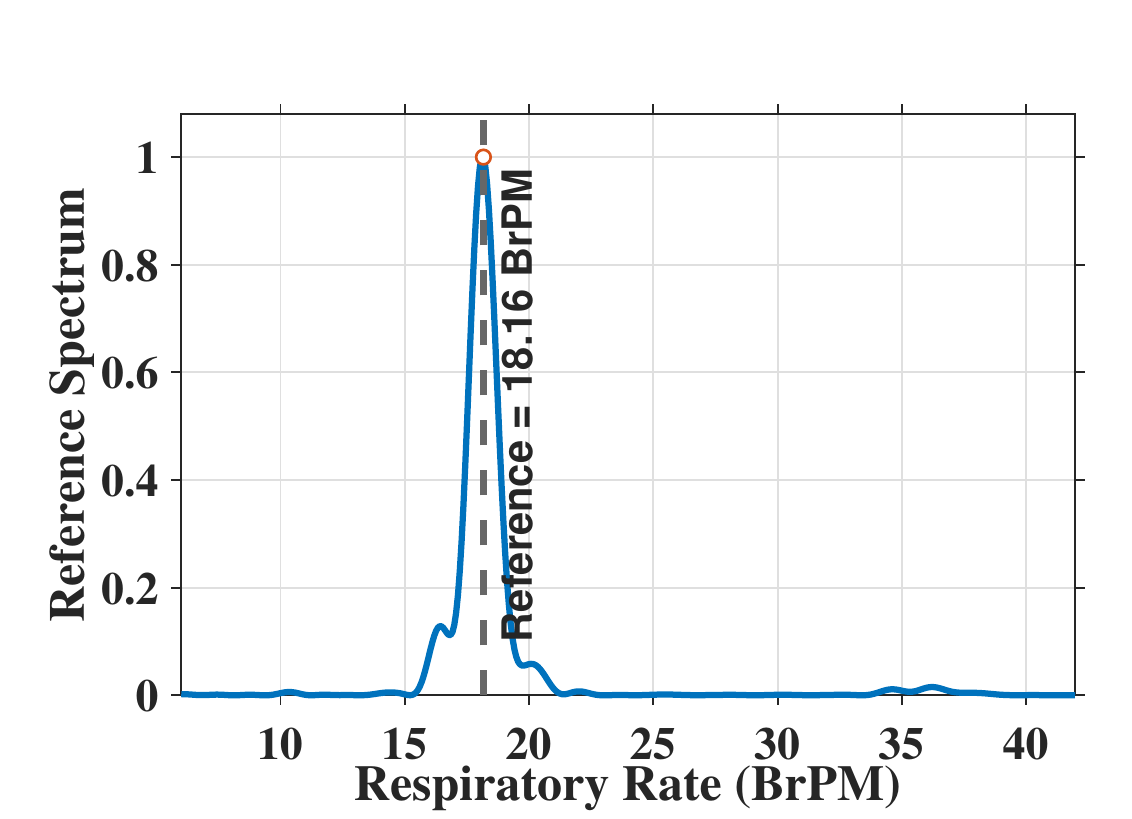}
\label{fig:baseline_record01_resp_ref}
}
\hfill
\subfloat[]{
\includegraphics[width=0.47\columnwidth]
{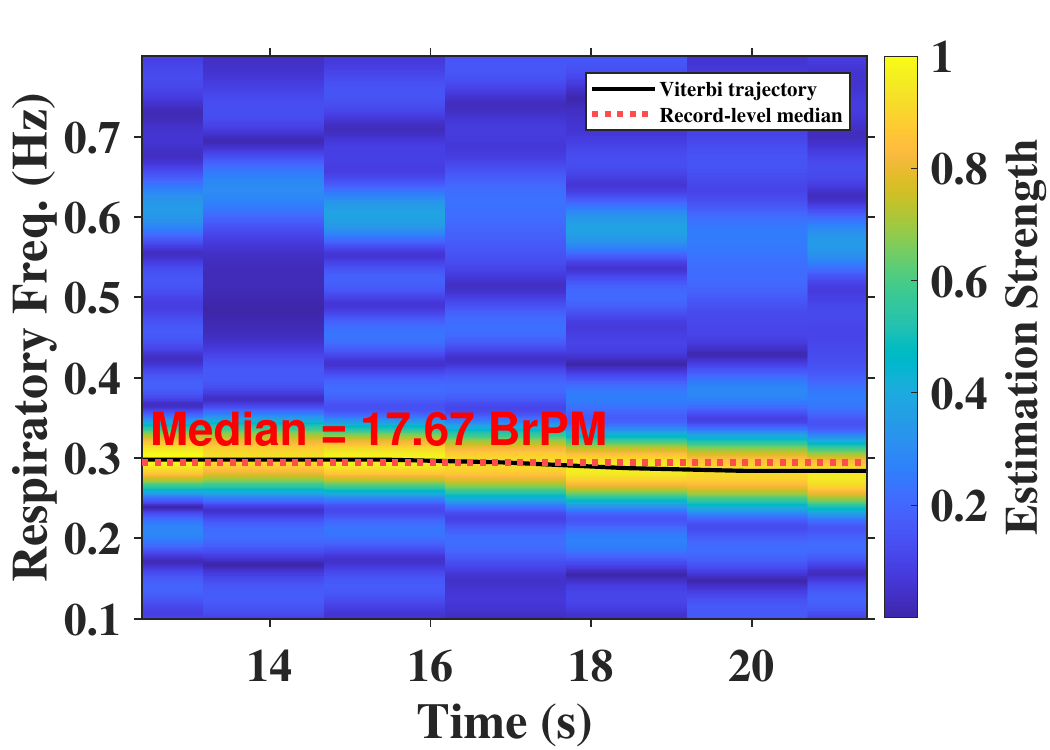}
\label{fig:baseline_record01_resp_radar}
}

\vspace{0.5em}

\subfloat[]{
\includegraphics[width=0.47\columnwidth]
{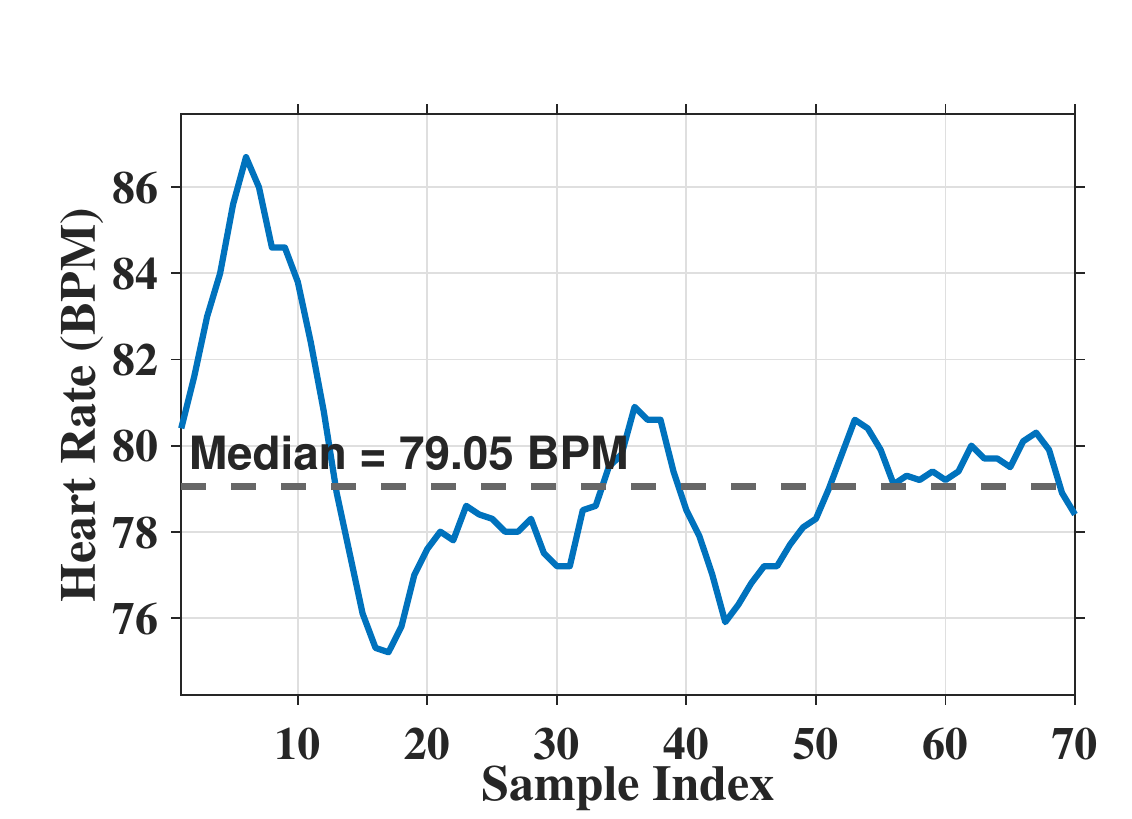}
\label{fig:baseline_record01_ecg}
}
\hfill
\subfloat[]{
\includegraphics[width=0.47\columnwidth]
{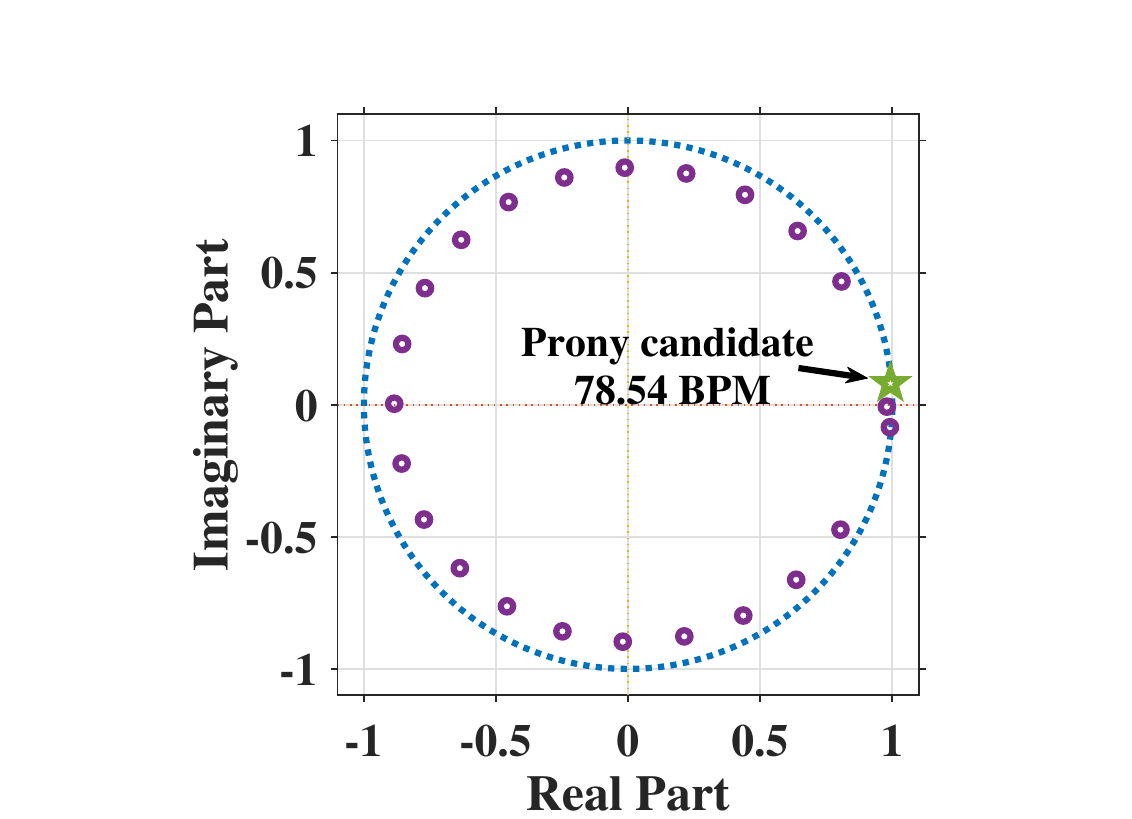}
\label{fig:baseline_record01_prony}
}

\caption{Representative baseline results for Record~01 at
$10~\mathrm{GHz}$: (a) spectrum of the signal from the respiration
belt, (b) radar respiratory frequency profile with the Viterbi
trajectory and median estimate for the record, (c) heart rate from the
ECG recording with its median value, and (d) Prony roots in the
$z$-plane with the selected baseline Prony candidate.}
\label{fig:baseline_record01}
\end{figure}

Fig.~\ref{fig:baseline_record01} illustrates the baseline processing
results for Record~01 at $10~\mathrm{GHz}$. The respiratory rate
estimated from the radar was $17.670~\mathrm{BrPM}$, compared with
$18.164~\mathrm{BrPM}$ from the respiration belt. The baseline Prony 
cardiac candidate was $78.536~\mathrm{BPM}$,
while the corresponding ECG reference was $79.050~\mathrm{BPM}$.

\subsection{End-Expiratory Apnea Positive Control}
\label{subsec:apnea_control}

A narrow bandpass filter can make even noise appear as a regular
oscillation. Therefore, a waveform that resembles cardiac activity only
after cardiac-band filtering does not by itself provide convincing
evidence that the corresponding component was present in the radar
measurements before filtering. For this reason, a separate
end-expiratory apnea acquisition was used as a positive control. We
examined the calibrated radar phase without cardiac-band filtering to
determine whether a common component was present before cardiac
filtering.

After a normal exhalation, the participant held their breath for
approximately $20~\mathrm{s}$. This substantially reduced periodic chest
motion caused by respiration during the analyzed interval. The $N_c=15$
calibrated phase sequences were examined without respiratory harmonic
notching or cardiac-band filtering. PCA was then applied directly to
these phase observations. Let
$\lambda_1\geq\lambda_2\geq\cdots\geq\lambda_{15}$ denote the PCA
eigenvalues. The normalized variance contribution of the $i$th
principal component is
\begin{equation}
\eta_i
=
\frac{\lambda_i}
{\displaystyle\sum_{j=1}^{15}\lambda_j}.
\label{eq:apnea_pca_variance}
\end{equation}

\begin{figure}[!t]
\centering
\includegraphics[width=0.82\linewidth]{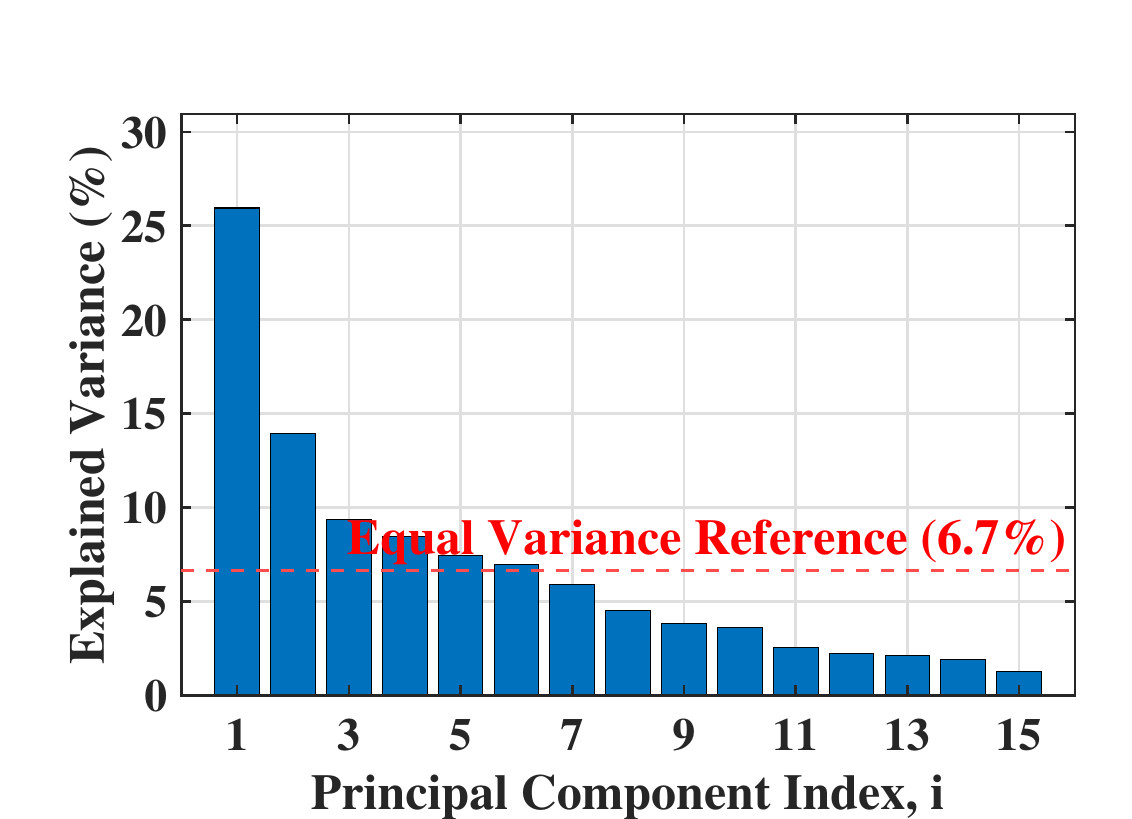}
\caption{Normalized variance contributions of the 15 principal components
during end-expiratory apnea. The dashed line denotes the equal-variance
reference of $1/15\approx6.7\%$.}
\label{fig:pca_apnea}
\end{figure}

Fig.~\ref{fig:pca_apnea} shows the normalized PCA eigenvalue
distribution. The first principal component accounted for $25.9\%$ of
the total variance. If the variance were distributed equally among the
15 components, each component would account for
$1/15\approx6.7\%$. The larger contribution of the first principal
component indicates that the phase observations contained a common
component. The $6.7\%$ value is used only as an equal variance
reference and not as a statistical threshold for rejecting white noise.

\begin{figure}[!t]
\centering
\subfloat[]{%
    \includegraphics[width=0.43\linewidth]{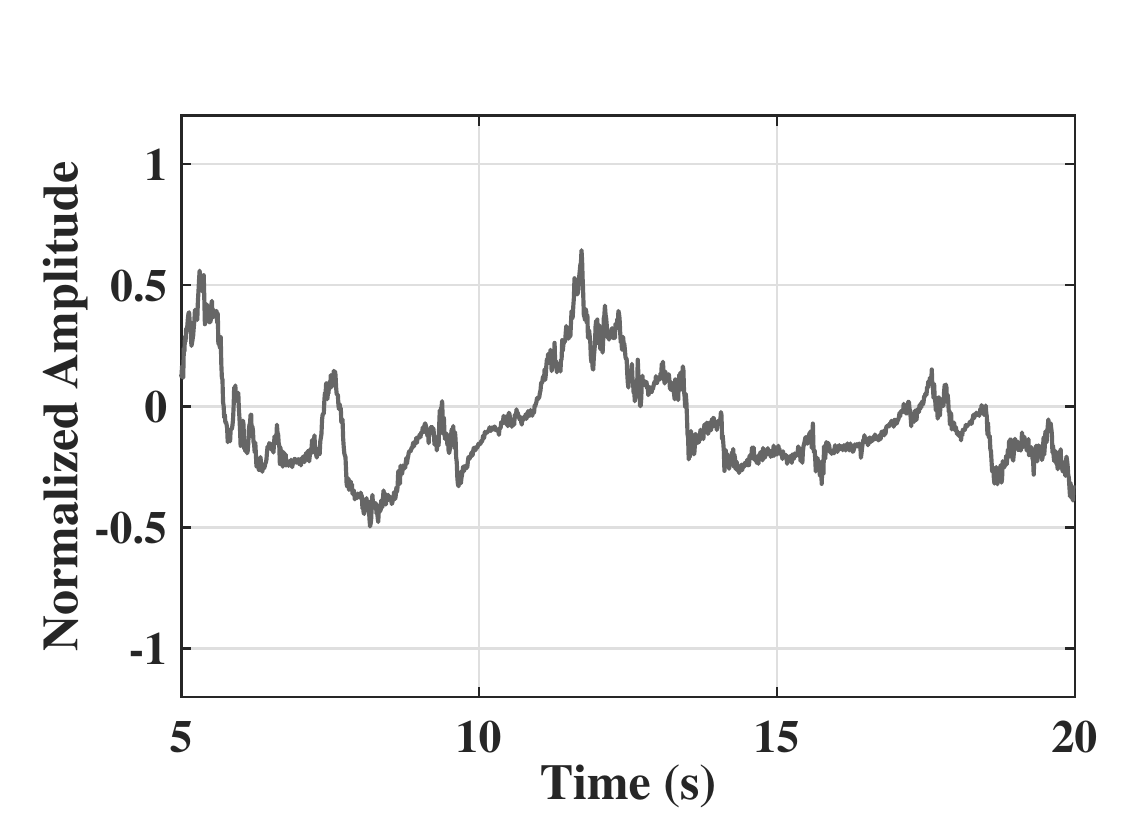}%
    \label{fig:Apnea_Raw}%
}\hfil
\subfloat[]{%
    \includegraphics[width=0.43\linewidth]{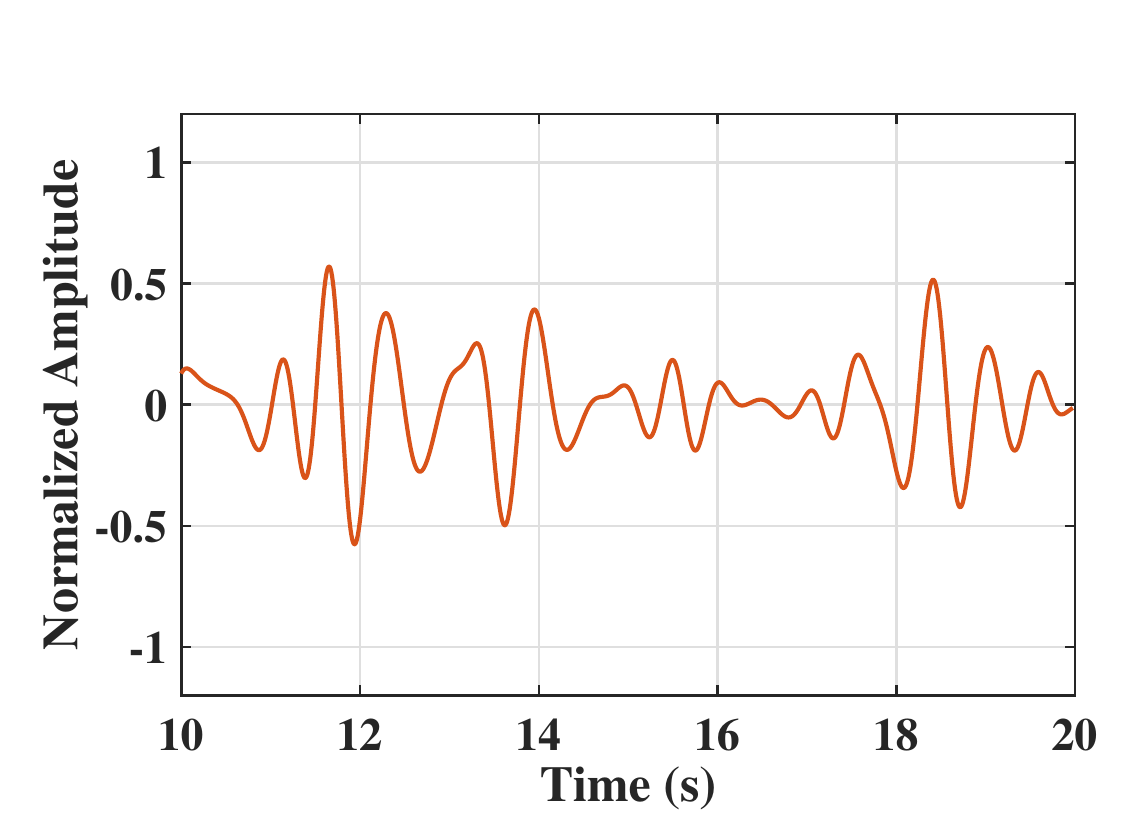}%
    \label{fig:Apnea_Waveform}%
}
\\
\vspace{1mm}
\subfloat[]{%
    \includegraphics[width=0.43\linewidth]{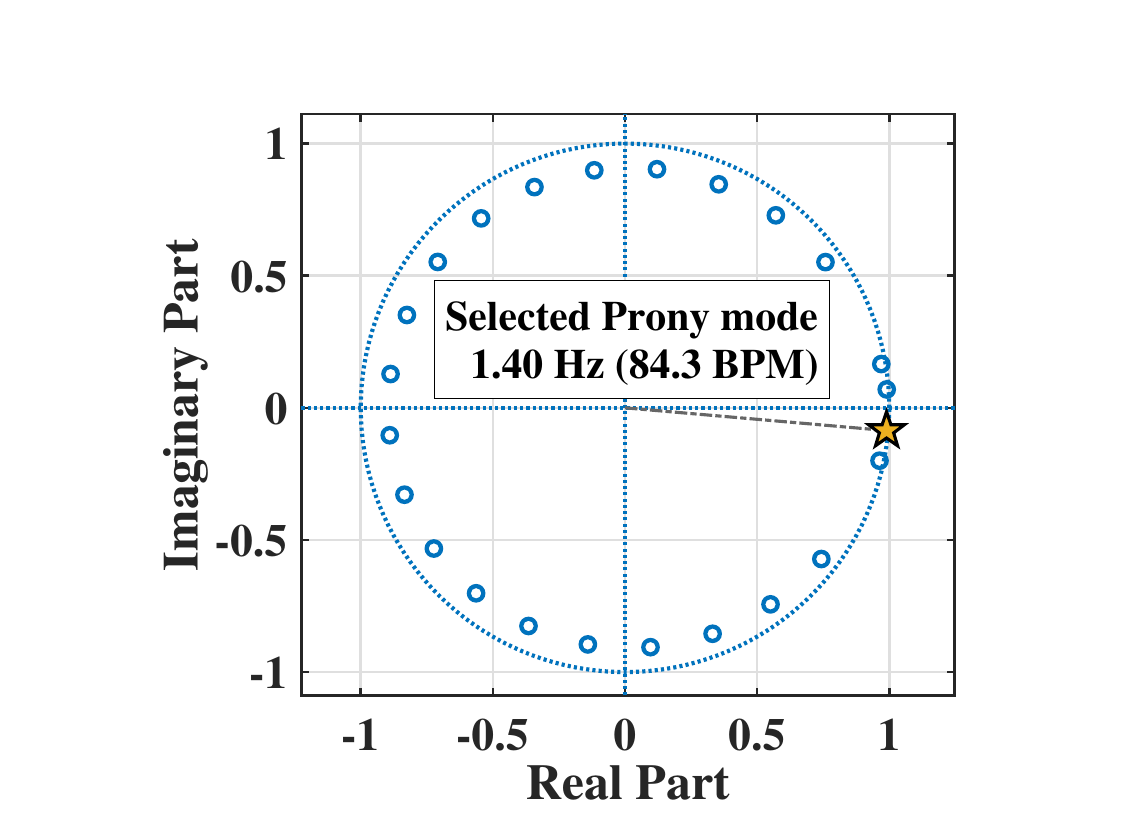}%
    \label{fig:Apnea_ZPlane}%
}\hfil
\subfloat[]{%
    \includegraphics[width=0.43\linewidth]{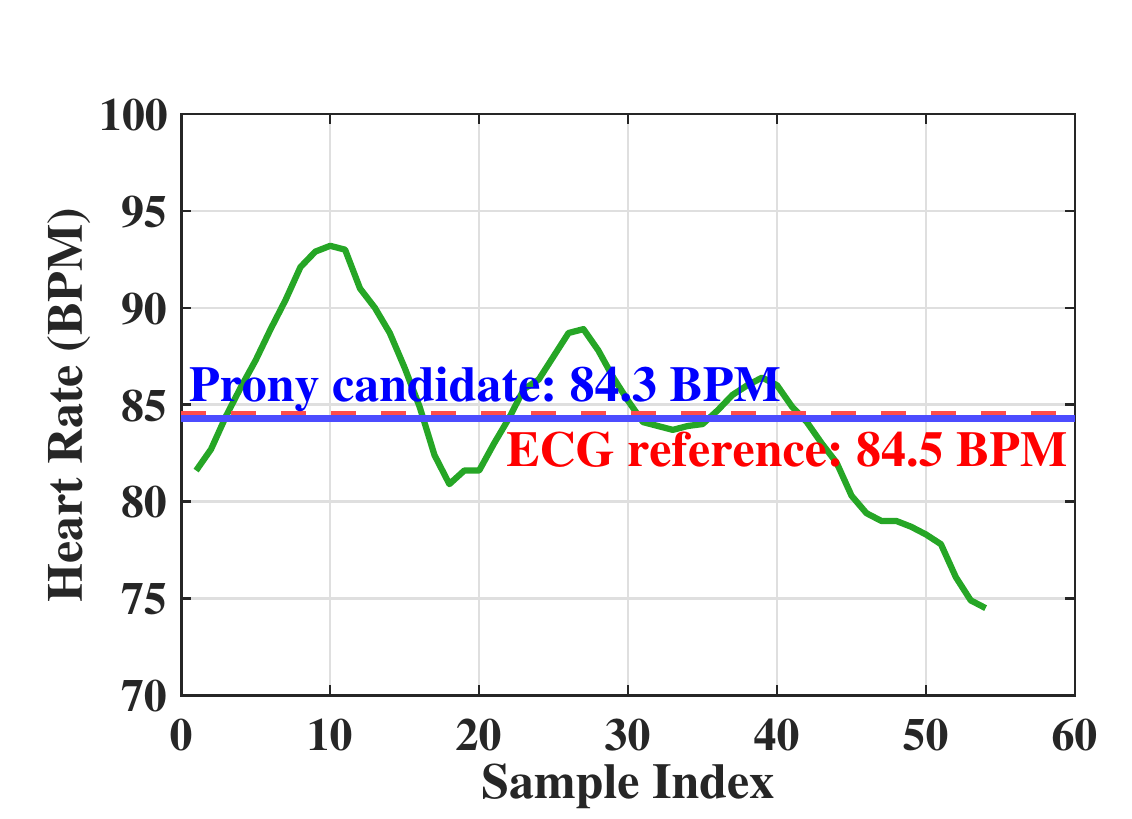}%
    \label{fig:Apnea_GT}%
}
\caption{End-expiratory apnea positive control:
(a) phase observation, (b) waveform after baseline cardiac processing,
(c) Prony roots in the $z$-plane, and (d) heart rate from the ECG recording.}
\label{fig:apnea_validation}
\end{figure}

Fig.~\ref{fig:apnea_validation} shows the cardiac results from
the apnea acquisition. The baseline cardiac processing yielded the
recovered waveform and a Prony cardiac candidate of
$84.3~\mathrm{BPM}$. The ECG reference for this acquisition was
approximately $84.5~\mathrm{BPM}$.
The radar cardiac estimate closely agrees with the ECG reference. This
supports the presence of cardiac-related motion in the radar measurements
before cardiac filtering.

\subsection{Respiratory Harmonic Ambiguity Assessment}
\label{subsec:harmonic_failure}

Respiratory motion is not purely sinusoidal and contains harmonic
components~\cite{11052672} that can extend into the frequency range
occupied by cardiac motion.
When a respiratory harmonic approaches the cardiac frequency, it can interfere
with cardiac estimation and make the physiological origin of a frequency
component ambiguous. Respiratory harmonic interference has been widely
reported in radar vital sign sensing, with mitigation methods including
high-resolution parametric estimation, adaptive multistage harmonic
suppression, cyclostationary harmonic suppression, and phase-difference
processing \cite{5247107,10579790,10945516}. This behavior is also
observed in our measurements, where the Prony cardiac candidate
approaches harmonics of the tracked respiratory frequency in several
records. Harmonic suppression or separation is outside the scope of
this work. We instead characterize the observed ambiguity in terms of
its frequency proximity, persistence over the record, and
time--frequency behavior.

To quantify this ambiguity, we compare the
record-level Prony cardiac candidate with the respiratory harmonics computed from the respiratory trajectory
tracked by Viterbi. Because the harmonic frequencies change with the respiratory
trajectory, this separation is evaluated in each slow-time analysis
window. For the $\ell$th window, the distance from the Prony candidate
to the nearest examined respiratory harmonic is defined as
\begin{equation}
\varepsilon_{\ell}
=
\min_{h\in\{3,\ldots,8\}}
\left|
\widehat{f}_{\mathrm{Prony}}
-
h\widehat{f}_{\mathrm{R},\ell}
\right|.
\label{eq:harmonic_distance}
\end{equation}
A window is counted as a collision when the nearest harmonic lies within
a fixed frequency guard of the Prony candidate,
\begin{equation}
\varepsilon_{\ell}
\leq
\varepsilon_{0},
\qquad
\varepsilon_{0}=0.08~\mathrm{Hz},
\label{eq:harmonic_collision}
\end{equation}
where the guard $\varepsilon_{0}$ is used only as an operational criterion
for geometric proximity. The $0.08~\mathrm{Hz}$ guard corresponds to $4.8~\mathrm{BPM}$,
close to the $5~\mathrm{BPM}$ tolerance used for cardiac error in
Section~\ref{subsec:baseline_performance}.
For a record containing $N_{\mathrm{win}}$ slow-time analysis windows,
the collision occupancy is defined as
\begin{equation}
\rho_{\mathrm{col}}
=
\frac{100}{N_{\mathrm{win}}}
\sum_{\ell=1}^{N_{\mathrm{win}}}
\mathbb{I}
\left(
\varepsilon_{\ell}\leq\varepsilon_{0}
\right),
\label{eq:collision_occupancy}
\end{equation}
where $\mathbb{I}(\cdot)$ denotes the indicator function. The minimum distance over all slow-time analysis windows in the record,
\begin{equation}
\varepsilon_{\min}
=
\min_{\ell}\varepsilon_{\ell},
\label{eq:minimum_harmonic_distance}
\end{equation}
is retained together with the harmonic order at which this minimum
occurs. 
Table~\ref{tab:harmonic_proximity} summarizes the proximity measures
for all 16 measurement records.

\begin{table}[t]
\caption{Proximity of the Prony candidate to the examined respiratory
harmonics across the 16 measurement records.}
\label{tab:harmonic_proximity}
\centering
\renewcommand{\arraystretch}{1.08}
\begin{tabular}{c c c c}
\hline
\textbf{Record} &
\textbf{Harmonic order at $\varepsilon_{\min}$} &
\textbf{$\varepsilon_{\min}$ (Hz)} &
\textbf{$\rho_{\mathrm{col}}$ (\%)} \\
\hline
01 & H5 & 0.1109 & 0.000 \\
02 & H4 & 0.0040 & 100.000 \\
03 & H4 & 0.0130 & 42.857 \\
04 & H5 & 0.0403 & 100.000 \\
05 & H4 & 0.0220 & 100.000 \\
06 & H5 & 0.1006 & 0.000 \\
07 & H4 & 0.0268 & 100.000 \\
08 & H5 & 0.0645 & 71.429 \\
09 & H5 & 0.1480 & 0.000 \\
10 & H5 & 0.0076 & 100.000 \\
11 & H4 & 0.0094 & 100.000 \\
12 & H5 & 0.0673 & 100.000 \\
13 & H5 & 0.0123 & 100.000 \\
14 & H5 & 0.0509 & 57.143 \\
15 & H5 & 0.1310 & 0.000 \\
16 & H4 & 0.1301 & 0.000 \\
\hline
\end{tabular}
\end{table}

\begin{figure}[!t]
\centering
\subfloat[]{%
    \includegraphics[width=0.49\linewidth]
    {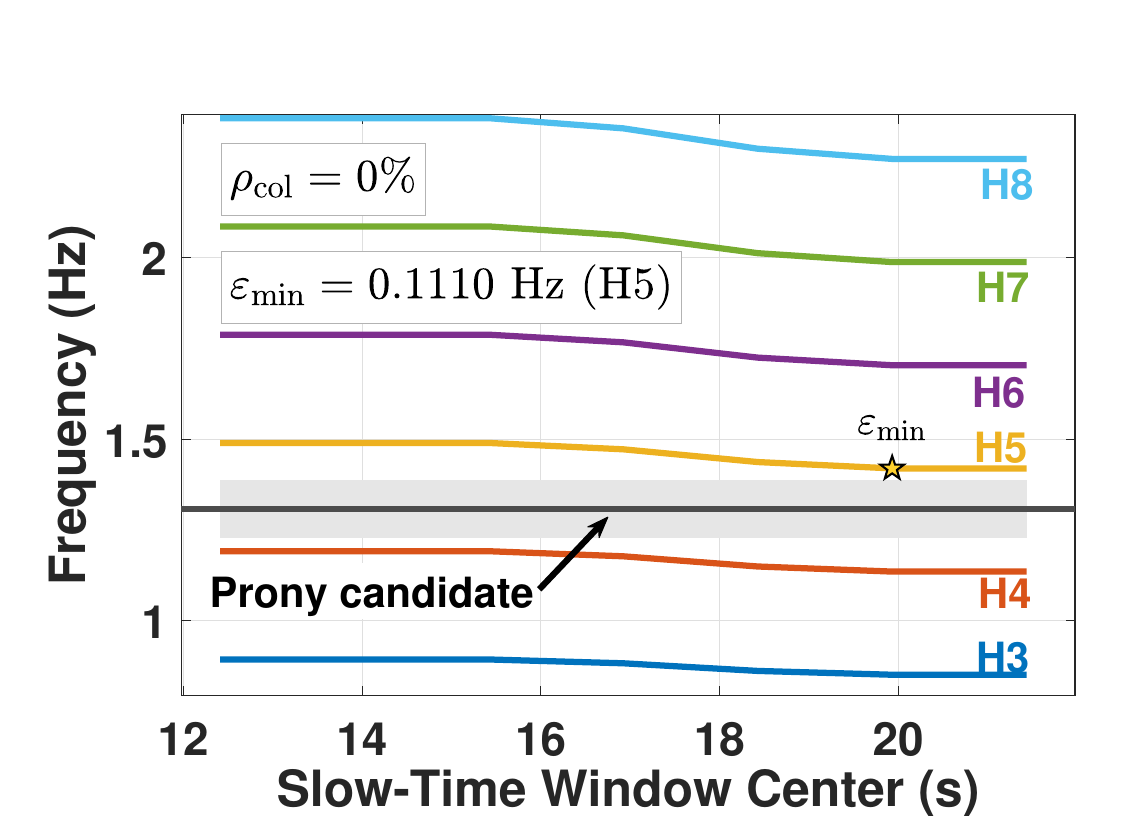}%
    \label{fig:harmonic_proximity_record01}%
}\hfil
\subfloat[]{%
    \includegraphics[width=0.49\linewidth]
    {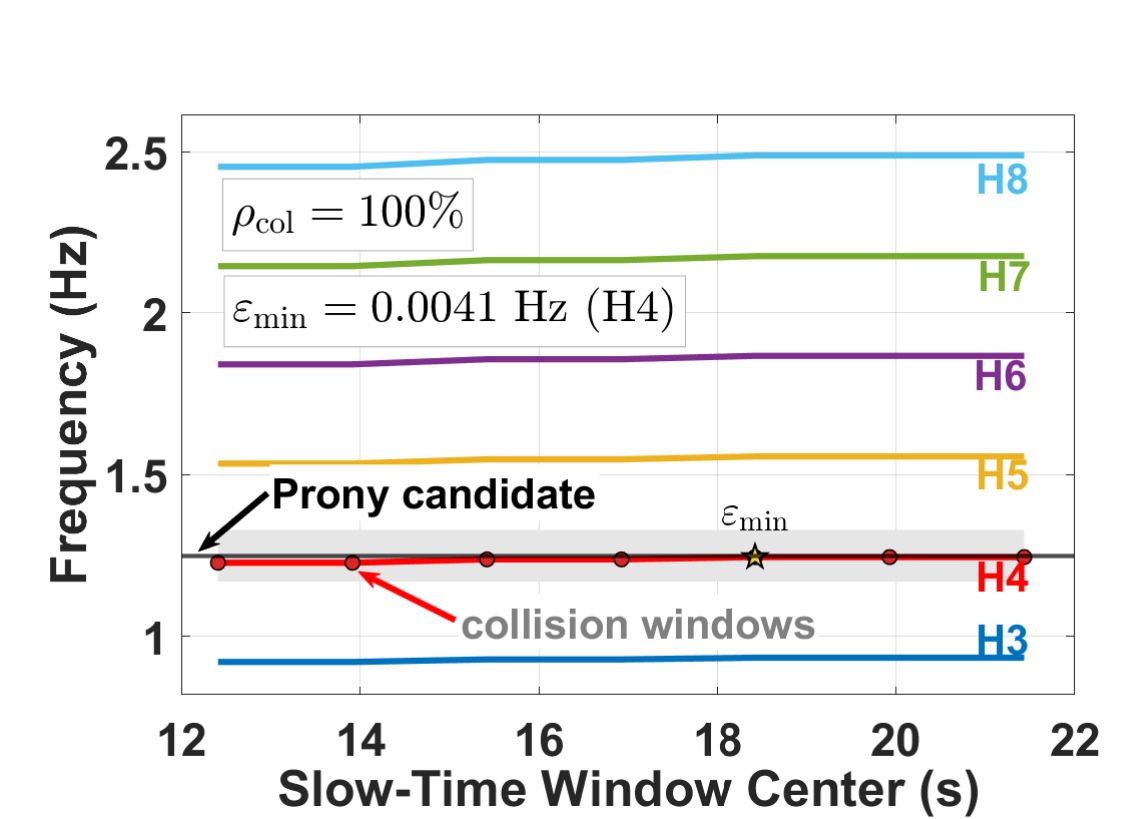}%
    \label{fig:harmonic_proximity_record02}%
}
\caption{Proximity of the Prony candidate to the examined respiratory
harmonics for (a) Record~01 and (b) Record~02. The gray band denotes the
$\pm0.08$~Hz frequency guard around the Prony candidate. The star marks
$\varepsilon_{\min}$, and red markers indicate collision windows.}
\label{fig:harmonic_proximity_examples}
\end{figure}

\begin{figure*}[!t]
    \centering

    \subfloat[\label{fig:rec03_cqu_tfr}]{
        \includegraphics[width=0.92\columnwidth]
        {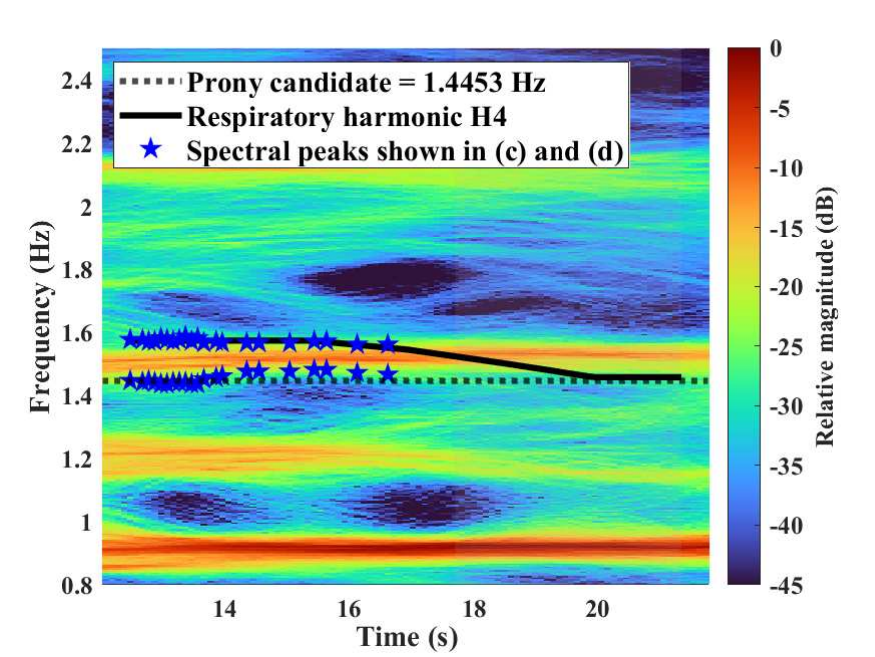}
    }
    \hspace{10mm}
    \subfloat[\label{fig:rec03_geometry}]{
        \includegraphics[width=0.92\columnwidth]
        {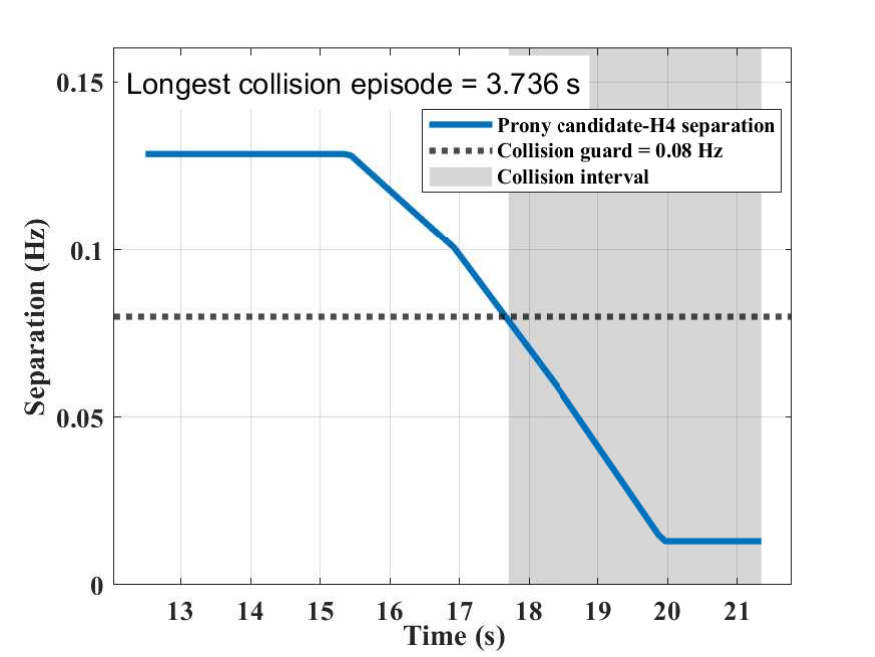}
    }

    \vspace{-1mm}

    \subfloat[\label{fig:rec03_pre_spectrum}]{
        \includegraphics[width=0.92\columnwidth]
        {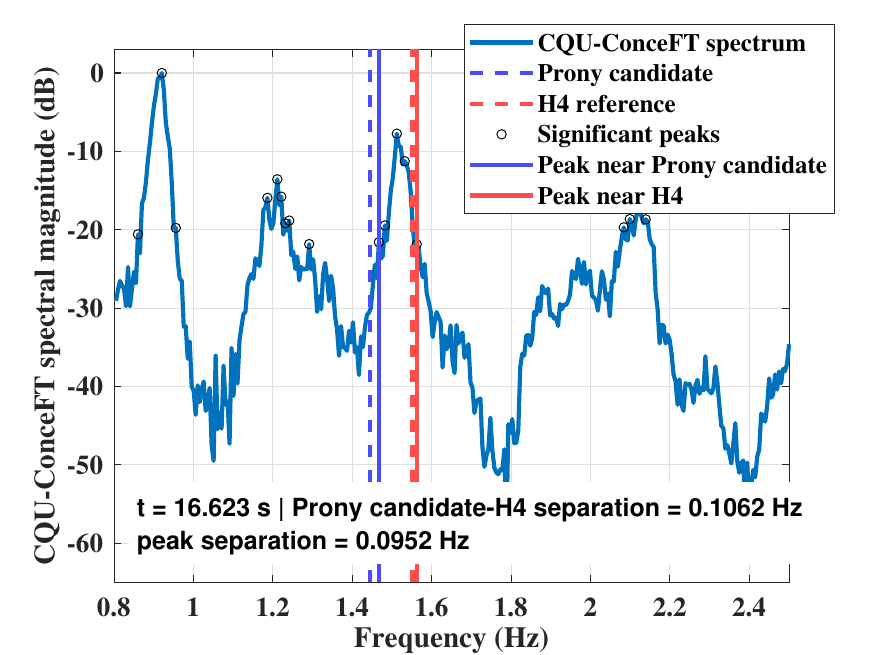}
    }
    \hspace{10mm}
    \subfloat[\label{fig:rec03_collision_spectrum}]{
        \includegraphics[width=0.92\columnwidth]
        {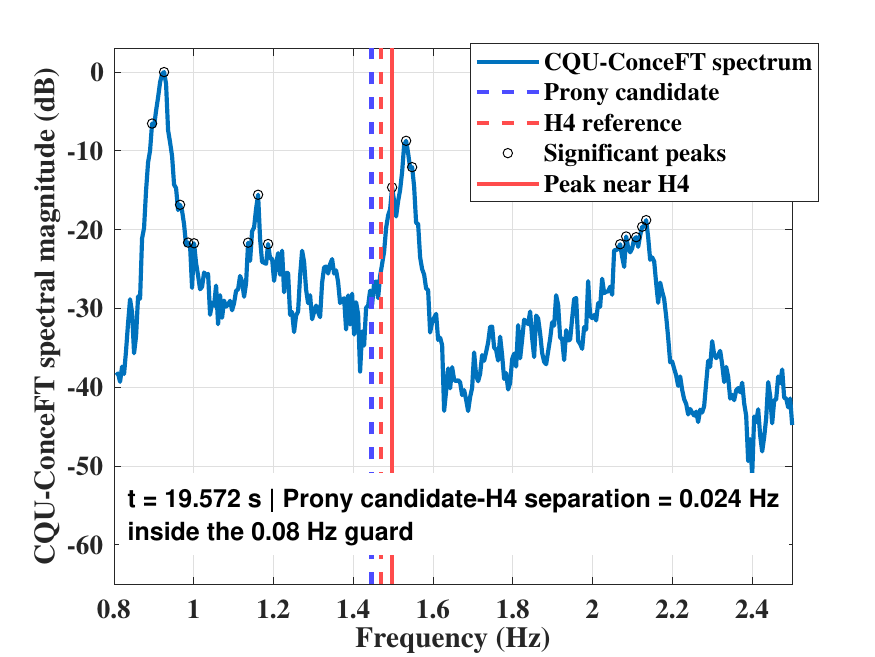}
    }

    \caption{Time--frequency and geometric analysis for Record~03:
    (a) CQU-ConceFT time-frequency representation, (b) candidate--H4 geometry relative to the
    $0.08~\mathrm{Hz}$ collision guard, (c) spectrum before the collision,
    and (d) spectrum during the collision.}
    \label{fig:record03_local_separability}
\end{figure*}

The proximity patterns vary substantially across the 16 records.
Five records exhibit no collision within the specified guard
($\rho_{\mathrm{col}}=0\%$), three exhibit partial collision occupancy,
and eight exhibit persistent proximity, with an examined respiratory
harmonic lying within the guard in every slow-time analysis window
($\rho_{\mathrm{col}}=100\%$).

The cardiac errors further illustrate the relationship between harmonic
proximity and estimation error. Four of the five
records with cardiac errors greater than $5~\mathrm{BPM}$ are Records~02,
04, 05, and 07, all of which have
$\rho_{\mathrm{col}}=100\%$. However, persistent proximity does not
always lead to a large error: Records~10--13 also have
$\rho_{\mathrm{col}}=100\%$ but cardiac errors no greater than
$5~\mathrm{BPM}$. Record~06 shows the opposite case, with a cardiac
error greater than $5~\mathrm{BPM}$ and
$\rho_{\mathrm{col}}=0\%$. Thus, proximity to a respiratory harmonic is
an important ambiguity in the measurements, but it does not by itself
explain the cardiac error in every record.

Fig.~\ref{fig:harmonic_proximity_examples} shows how the proximity
between the Prony candidate and the respiratory harmonics changes over
time for Records~01 and~02. In Record~01, none of the examined
respiratory harmonics enters the $\pm\varepsilon_0$ guard around the
Prony candidate, resulting in $\rho_{\mathrm{col}}=0\%$. In Record~02,
the fourth harmonic (H4) trajectory remains within the guard in every
slow-time analysis window, leading to $\rho_{\mathrm{col}}=100\%$.

The geometric measures show when the Prony candidate approaches a
respiratory harmonic, but not whether separate spectral components remain
visible as their frequencies approach each other. We examine this question
in the records with partial collision occupancy using time--frequency
analysis.

Concentration of Frequency and Time (ConceFT) was introduced for
time--frequency analysis of noisy multicomponent signals with
time-varying amplitudes and instantaneous frequencies
\cite{10.1098/rsta.2015.0193}, and has been applied to physiological
signals, including heart rate variability (HRV) and QT variability
from ECG recordings \cite{7445203,Wu2018HRVQT}. The complex
quasi-uniform (CQU) variant, CQU-ConceFT, replaces the random
projections of ConceFT with deterministic complex spherical designs
\cite{SOURISSEAU2021137}, and has been demonstrated on physiological
data for the analysis of time-varying spindle structure in N2-stage
electroencephalogram (EEG) recordings \cite{SOURISSEAU2021137}.

In this work, CQU-ConceFT is used only as a diagnostic for the
partial-collision records. It examines whether separate spectral
components remain visible before and/or after the collision and how their
time--frequency structure changes as the respiratory harmonic approaches
the Prony candidate. This evidence supplements the geometric proximity
analysis.

For the radar measurements, CQU-ConceFT is applied to
the same phase observation obtained after PCA fusion in the baseline
cardiac processing. The
fixed H1--H5 respiratory harmonic notch filters are omitted so that the
harmonic structure remains present in the observation. The signal is
restricted to the $0.8$--$2.5~\mathrm{Hz}$ cardiac frequency range
before the time--frequency analysis. CQU-ConceFT is implemented
following the formulation of Sourisseau \emph{et al.}
\cite{SOURISSEAU2021137}; implementation details are provided in that
reference.

Record~03 provides a representative example of partial collision, with
$\rho_{\mathrm{col}}=42.857\%$ and H4 as the respiratory harmonic
associated with the minimum separation. The CQU-ConceFT representation
for this record is shown in Fig.~\ref{fig:rec03_cqu_tfr}. Before the
collision, at $t=16.623~\mathrm{s}$, the separation between the Prony
candidate and H4 is $0.1062~\mathrm{Hz}$, as shown in
Fig.~\ref{fig:rec03_geometry}. At the same time, the spectrum in
Fig.~\ref{fig:rec03_pre_spectrum} shows two nearby spectral peaks at
approximately $1.4670$ and $1.5622~\mathrm{Hz}$, corresponding to a
separation of $0.0952~\mathrm{Hz}$. The two spectral structures are
therefore distinguishable before the collision, consistent with the
geometric separation outside the collision guard.

Later, at $t=19.572~\mathrm{s}$, the candidate--H4 separation decreases
to $0.0239~\mathrm{Hz}$, placing the two frequencies within the
$0.08~\mathrm{Hz}$ collision guard, as shown in
Fig.~\ref{fig:rec03_geometry}. The corresponding spectrum during the 
collision is shown in Fig.~\ref{fig:rec03_collision_spectrum}.
Thus, up to the onset of the collision, separate spectral components
remain distinguishable in the time--frequency representation, and the
geometric criterion does not indicate candidate--harmonic ambiguity.
Once H4 enters the collision region, the geometric criterion identifies
candidate--harmonic ambiguity.

Records~03, 08, and~14 are the three records with partial collision
occupancy. In all three records, separate spectral components are
visible outside the collision region, while the corresponding
record-level cardiac estimates remain within $5~\mathrm{BPM}$ of the
ECG references. These observations illustrate both the record-level
agreement with ECG in the three partial-collision cases and the
time-varying respiratory-harmonic ambiguity captured by the geometric
criterion.

\section{Conclusion}
\label{sec:conclusion}
This work presented, to the best of our knowledge, the first experimental demonstration of contactless vital-sign sensing in the FR3 band.
This demonstration was enabled by VitalFR3, a programmable FMCW measurement platform that integrates direct sampling with an Ettus USRP X440 and a SignalCraft SC2470 FR3 upconverter.
The results
showed that the complete transmit-receive chain should be treated as a
single measurement instrument and characterized in amplitude, phase, and
timing before physiological signal processing. Once this measurement reference is
established, the calibrated complex I/Q measurements can be used for
fast-time localization of a persistent target response and to extract the
respiratory and cardiac motion across multiple FR3 center frequencies.

Our measurements and processing algorithms demonstrated excellent agreement with the 
reference respiratory rates, while the fine cardiac estimation was more challenging
in some measurements due to the dominance of the respiratory signal. 
An apnea measurement study provided additional
evidence that cardiac-related motion could be better recovered when respiratory motion was reduced.
Under normal breathing, however, the measurements revealed that respiratory harmonics can enter the cardiac frequency
range, making frequency alone insufficient to determine the physiological
origin of a component. Future work should extend the validation to more
subjects and measurement conditions, and investigate methods that use
additional signal information to distinguish cardiac motion from respiratory
harmonics when their frequencies approach or overlap.

\section*{Acknowledgment}
The authors would like to thank N. Pandeya and M. Dickens of Ettus Research
for their technical guidance and support in configuring the
USRP X440 testbed; the SignalCraft Technologies team, particularly
T. Wilson, for technical support
related to the SC2470 FR3 up/down converter; and
M. Majidimosleh of the iRadio Lab, University of Calgary, for discussions
and technical advice on RF aspects of the experimental setup.

\bibliographystyle{IEEEtran}
\bibliography{mybibfile_IEEE_style}

\end{document}